\documentclass[twoside,twocolumn,9pt]{article}
\usepackage{extsizes}
\usepackage[super,sort&compress,comma]{natbib} 
\usepackage[version=3]{mhchem}
\usepackage[left=1.5cm, right=1.5cm, top=1.785cm, bottom=2.0cm]{geometry}
\usepackage{balance}
\usepackage{mathptmx}
\usepackage{sectsty}
\usepackage{graphicx} 
\usepackage{lastpage}
\usepackage[format=plain,justification=justified,singlelinecheck=false,font={stretch=1.125,small,sf},labelfont=bf,labelsep=space]{caption}
\usepackage{float}
\usepackage{fancyhdr}
\usepackage{fnpos}
\usepackage[english]{babel}
\addto{\captionsenglish}{%
  
}
\usepackage{array}
\usepackage{droidsans}
\usepackage{charter}
\usepackage[T1]{fontenc}
\usepackage[usenames,dvipsnames]{xcolor}
\usepackage{setspace}
\usepackage[compact]{titlesec}
\usepackage{hyperref}
\usepackage{textgreek}
\usepackage{siunitx}
\usepackage{amsmath,amssymb}

\usepackage[normalem]{ulem}

\usepackage{epstopdf}

\definecolor{cream}{RGB}{222,217,201}

\begin{document}

\pagestyle{fancy}
\thispagestyle{plain}
\fancypagestyle{plain}{
\renewcommand{\headrulewidth}{0pt}
}

\makeFNbottom
\makeatletter
\renewcommand\LARGE{\@setfontsize\LARGE{15pt}{17}}
\renewcommand\Large{\@setfontsize\Large{12pt}{14}}
\renewcommand\large{\@setfontsize\large{10pt}{12}}
\renewcommand\footnotesize{\@setfontsize\footnotesize{7pt}{10}}
\makeatother

\renewcommand{\thefootnote}{\fnsymbol{footnote}}
\renewcommand\footnoterule{\vspace*{1pt}%
\color{cream}\hrule width 3.5in height 0.4pt \color{black}\vspace*{5pt}} 
\setcounter{secnumdepth}{5}

\makeatletter 
\renewcommand\@biblabel[1]{#1}            
\renewcommand\@makefntext[1]%
{\noindent\makebox[0pt][r]{\@thefnmark\,}#1}
\makeatother 
\renewcommand{\figurename}{\small{Fig.}~}
\sectionfont{\sffamily\Large}
\subsectionfont{\normalsize}
\subsubsectionfont{\bf}
\setstretch{1.125} 
\setlength{\skip\footins}{0.8cm}
\setlength{\footnotesep}{0.25cm}
\setlength{\jot}{10pt}
\titlespacing*{\section}{0pt}{4pt}{4pt}
\titlespacing*{\subsection}{0pt}{15pt}{1pt}

\fancyfoot{}
\fancyfoot[LO,RE]{\vspace{-7.1pt}\includegraphics[height=9pt]{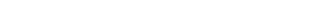}}
\fancyfoot[CO]{\vspace{-7.1pt}\hspace{13.2cm}\includegraphics{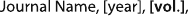}}
\fancyfoot[CE]{\vspace{-7.2pt}\hspace{-14.2cm}\includegraphics{head_foot/RF}}
\fancyfoot[RO]{\footnotesize{\sffamily{1--\pageref{LastPage} ~\textbar  \hspace{2pt}\thepage}}}
\fancyfoot[LE]{\footnotesize{\sffamily{\thepage~\textbar\hspace{3.45cm} 1--\pageref{LastPage}}}}
\fancyhead{}
\renewcommand{\headrulewidth}{0pt} 
\renewcommand{\footrulewidth}{0pt}
\setlength{\arrayrulewidth}{1pt}
\setlength{\columnsep}{6.5mm}
\setlength\bibsep{1pt}

\makeatletter 
\newlength{\figrulesep} 
\setlength{\figrulesep}{0.5\textfloatsep} 

\newcommand{\topfigrule}{\vspace*{-1pt}%
\noindent{\color{cream}\rule[-\figrulesep]{\columnwidth}{1.5pt}} }

\newcommand{\botfigrule}{\vspace*{-2pt}%
\noindent{\color{cream}\rule[\figrulesep]{\columnwidth}{1.5pt}} }

\newcommand{\dblfigrule}{\vspace*{-1pt}%
\noindent{\color{cream}\rule[-\figrulesep]{\textwidth}{1.5pt}} }

\makeatother

\twocolumn[
  \begin{@twocolumnfalse}
{\includegraphics[height=30pt]{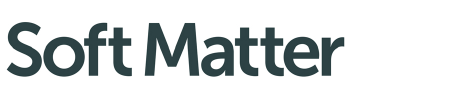}\hfill\raisebox{0pt}[0pt][0pt]{\includegraphics[height=55pt]{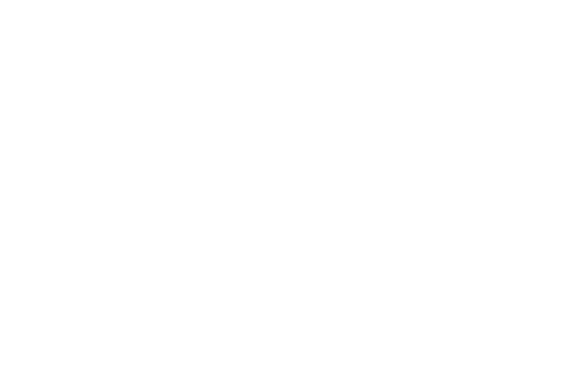}}\\[1ex]
\includegraphics[width=18.5cm]{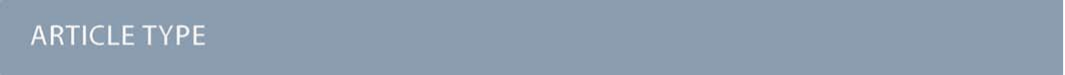}}\par
\vspace{1em}
\sffamily
\begin{tabular}{m{4.5cm} p{13.5cm} }

\includegraphics{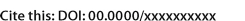} & \noindent\LARGE{\textbf{Sizing multimodal suspensions with differential dynamic microscopy}} \\
\vspace{0.3cm} & \vspace{0.3cm} \\

 & \noindent\large{Joe J Bradley,$^{\ast}$\textit{$^{a}$} Vincent A Martinez,\textit{$^{a}$}, Jochen Arlt,\textit{$^{a}$} John R Royer\textit{$^{a}$} and Wilson C K Poon\textit{$^{a}$}} \\

\includegraphics{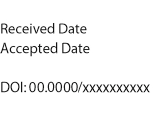} & \noindent\normalsize{Differential dynamic microscopy (DDM) can be used to extract the mean particle size from videos of suspensions. However, many suspensions have multimodal particle size distributions, for which a single `mean' is not a sufficient description. After clarifying how different particle sizes contribute to the signal in DDM, we show that standard DDM analysis can extract the mean sizes of two populations in a bimodal suspension given prior knowledge of the sample's bimodality. Further, the use of the CONTIN algorithm obviates the need for such prior knowledge. Finally, we show that by selectively analysing portions of the DDM images, we can size a trimodal suspension where the large particles would otherwise dominate the signal, again without prior knowledge of the trimodality.}

\end{tabular}

 \end{@twocolumnfalse} \vspace{0.6cm}

  ]

\renewcommand*\rmdefault{bch}\normalfont\upshape
\rmfamily
\section*{}
\vspace{-1cm}


\footnotetext{\textit{$^{a}$~School of Physics \& Astronomy, The University of Edinburgh, Peter Guthrie Tait Road, Edinburgh EH9 3FD, United Kingdom. E-mail: w.poon@ed.ac.uk}}




\newcommand\micron{~\SI{}{\micro\metre}}
\newcommand\x{$\times$}
\newcommand\degrees{$^{\circ}$}

\section{Introduction}\label{sec:Intro}

Particle sizing is important across many industrial sectors. A modern text\cite{merkus2009} lists seven categories of methods: microscopy, sieving, electrozoning, laser diffraction (= static light scattering, SLS), ultrasound extinction, sedimentation, and dynamic light scattering (DLS). Some of these measure particles one at a time (microscopy, electrozoning), others deal with collections of particles {\it en masse}. Many are optically based (various light microscopies, SLS, DLS). 

These methods are calibrated against quasi-monodisperse spherical particles, where the polydispersity, defined as the standard deviation of the particle size distribution (PSD) normalised by the mean, is typically $\lesssim 10\%$, and can even be $\lesssim 2\%$.\cite{Vrij1992} The sizing of such particles poses few problems; reporting simply a mean diameter and a polydispersity generally suffices.  

While quasi-monodisperse spheres find use in research and specialised applications, most real-life suspensions are significantly polydisperse, often with strongly-peaked, multimodal PSDs. Examples include raw and UHT milk, with a bimodal mixture of large fat droplets and smaller casein micelles,\cite{milk2014} sunflower tahini with a reported trimodal PSD,\cite{tahini2014} and chocolate, where the PSD shifts from trimodal to bimodal as refining proceeds.\cite{Tan2018} Multimodal PSDs can result from aggregation, for example nanoparticles used for biomedical applications often develop a second population of large agglomerates when dispersed in a physiological buffer.\cite{bionano2016}

Reporting a mean and polydispersity for a multimodal suspension is almost meaningless; ideally, one wants to capture the full PSD. In practice, it is often difficult to detect multimodality in the first place, let alone obtain mean sizes for each population. 

Direct imaging is perhaps the `gold standard' of sizing. However, the PSD must be built up particle by particle, requiring a large number, $N$, to accumulate statistics, with the relative uncertainty dropping only weakly as $N^{-\frac{1}{2}}$. Moreover, it is difficult to guarantee representative sampling, and preparation (e.g. drying for electron microscopy) can affect the particles. 

Scattering allows better statistical averaging, because the scattering volume typically contains many more particles than can be practically imaged. However, analysis requires inverting a Laplace transform, where the unknown PSD occurs under an integral sign, so that a unique solution does not exist and the problem is notoriously noise sensitive.\cite{Twomey1977} Nevertheless, various scattering methods, especially SLS and DLS, are popular, with many available commercial instruments and sophisticated analysis software (e.g.~CONTIN for DLS). Impressive answers can be obtained if some sample details are known. For example, SLS has been applied to a multimodal suspension with 5 populations varying in size over several orders of magnitude,\cite{LS_Multimodal} but requires the particles' refractive indices which are not trivial to obtain.

Differential dynamic microscopy (DDM) is a technique for high-throughput sizing in which the intermediate scattering function~(ISF), familiar from DLS, is obtained from images without the need to resolve the particles.\cite{Cerbino2008DDM} Since DDM and DLS both access the ISF, there is significant overlap in data analysis. However, DDM offers certain advantages, such as the ability to cope with significant turbidity.\cite{DDMTurbid} Here we show that DDM is well-suited for sizing multimodal suspensions because it probes spatial fluctuations at very low wave vector, $k$, even $\lesssim \SI{0.5}{\per\micro\meter}$, by imaging large fields of view.  Equivalent scattering angles of $\lesssim \SI{2}{\degree}$ in DLS require complex instrumentation,\cite{CipellettiIntermediateDLS, CipellettiLowAngleDLS} and are seldom attempted. 

\begin{figure}[t]
\centering
\includegraphics[width=\linewidth]{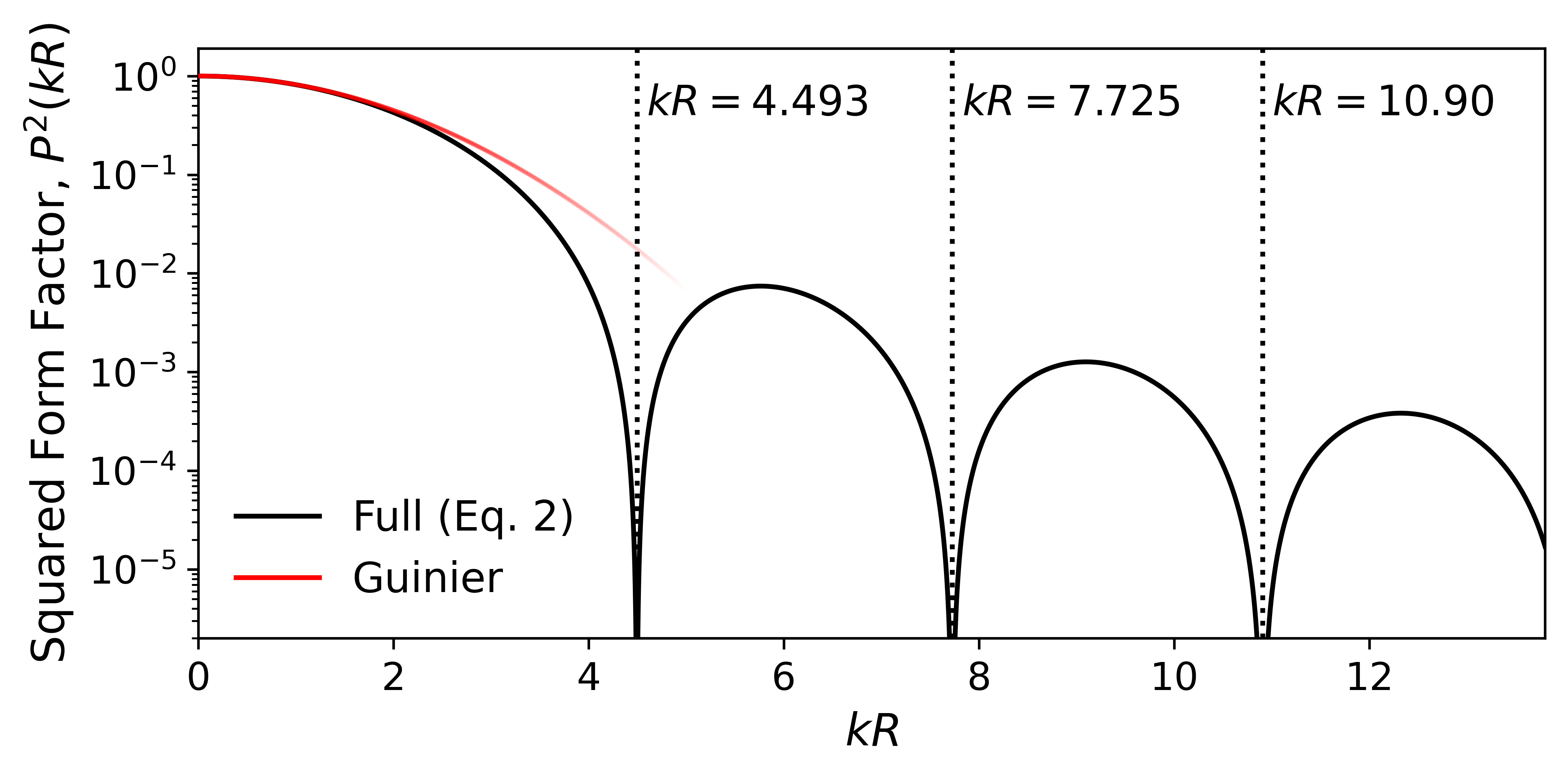}
  \caption{Theoretical squared form factor, $P^2(kR)$, for a monodisperse sphere of radius $R$ as a function of the scattering vector or Fourier component $k$ non-dimensionalised by the radius, $kR$, Eq.~\ref{eq:formfac}, with minima positions given. The red curve is the Guinier approximation.}
  \label{fgr:form_factor}
\end{figure}

In SLS and DLS, the electric field scattered by a single homogeneous sphere of radius $R$ at scattering vector $k$ is given by\cite{BombannesPusey}
\begin{equation}
b(k) = \left[ \frac{4}{3}\pi R^3 \right]\Delta n(k) P(k), \label{eq:bk}
\end{equation}
where $\Delta n(k)$ is the difference in refractive index between the particle and the solvent, and \begin{equation}
    P(kR) =  3[\sin(kR) - (kR)\cos(kR)]/(kR)^3,
    \label{eq:formfac}
\end{equation}
is the form factor with $P(kR \to 0) \to 1$, typically accessible experimentally only as the squared form factor $P^2(kR)$, Fig.~\ref{fgr:form_factor}.
$P(kR)$ displays successive zeros, with the first at $k_0R = 4.493$. Two consequences follow from Eq.~\ref{eq:formfac}. First, in the dilute limit, the DLS signal scales as $N b^2(k)$ for $N$ particles in the scattering volume,\cite{BombannesPusey} so at low $k$ as $NR^6 \sim \phi R^3$, where $\phi$ is the particle volume fraction. Secondly, particles with $R \approx 4.493/k_0$ in a polydisperse sample contribute little signal at scattering angles around the minimum. This effect can be used to measure low polydispersities accurately in a multi-angle experiment,\cite{PuseyPoly} but may generate large errors in commercial single-angle instruments.\cite{DLSPitfalls}

The DDM signal is also dependent on $P(kR)$.\cite{aniso_DDM} However, its first minimum has little effect, because for all Brownian suspensions DDM can operate with $kR \ll 1$ where $P(kR) \to 1$. So, Safari \textit{et al.}\cite{SafariBidisperseDDM} were able to use DDM to size a bidisperse suspension with a 1:20 particle size ratio and up to 3\% by volume of the large particles, where in all but one case, DLS fails to size the minority (large) species. However, these authors explicitly input the bimodality of their suspensions to their DDM analysis. 

In this work, we demonstrate the use of DDM to size a bidisperse suspension with a significantly smaller size ratio of 1:4.6 {\it without} assuming bimodality in the analysis, and probe the technique's efficacy when the number ratio of the species is systematically changed. Furthermore, we show how to extend the limits of applicability of DDM further by selecting regions of interest in our image sequences for analysis. The method is demonstrated by sizing a trimodal system in which the signal from the largest particles dominates, without assuming trimodality {\it a priori}. 

Below, we first present expressions for fitting DDM results to data from polydisperse suspensions, explaining how signal contribution scales with particle size. Next, we explain our experimental and data fitting methods. After validating our predicted signal vs.~size scaling, we demonstrate the application of DDM to bi- and tri-modal dispersions, concluding with a recommended protocol for sizing multimodal suspensions with DDM. 

\section{DDM for Polydisperse Suspensions} \label{sec:PolydisperseDDM}

The original implementation of DDM\cite{Cerbino2008DDM,Giavazzi2009DDM} used partially coherent illumination (see further Appendix~\ref{appendix:ScatteringDDM}). Since we perform bright-field imaging with a condenser of numerical aperture $\approx 0.5$, our illumination is incoherent. We have previously shown\cite{aniso_DDM} that for $N$ identical particles in an incoherent image, the differential image correlation function (DICF), $g(k,\tau)$, which is the squared Fourier transform of the difference between an image at time $\tau$ and a reference image at time zero, is related to the ISF, $f(k,\tau)$ by:
\begin{eqnarray}
    g(k, \tau) & = & A(k)\left[1 - f(k, \tau)\right] + B(k), \label{eq:DICF}\\
    A(k) & = & 2Na^2(k)S(k). \label{eq:Ak}
\end{eqnarray}
In this expression, which is the same as that derived in the partially-coherent limit,\cite{Cerbino2008DDM,Giavazzi2009DDM} $B(k)$ is the system's noise spectrum. The DDM signal\footnote[1]{There is no literature consensus on this nomenclature, descriptions of $A(k)$ include DDM `signal'\cite{Multiscale_Review}, `signal amplitude'\cite{SafariBidisperseDDM}, and `signal prefactor' and `signal term'\cite{Fluoresence_DDM}. Crucially, this should not be confused with the camera signal.}, $A(k) \propto a^2(k)$, the contribution from a single particle, and $\propto S(k)$, the particles' structure factor. Note, in passing, that $k$ here is a Fourier component of density fluctuations and not a scattering vector.\footnote[2]{To show that the signal measured at scattering vector $k$ in DLS in fact characterises density fluctuations with that wave vector requires considerable analysis.\cite{BombannesPusey}}

In a monodisperse suspension of non-interacting spherical particles of radius $R$, the ISF is $f(k,\tau) = \exp{[-Dk^2\tau]}$, with the diffusivity $D = k_BT/6\pi\eta R$ in a suspending medium of viscosity $\eta$ at temperature $T$ (and $k_B$ is Boltzmann's constant). So, fitting the measured $g(k,\tau)$ to Eq.~\ref{eq:DICF} returns $D$ and therefore $R$. Appendix~\ref{appendix:DDM_Theory} shows that these results generalise naturally to a suspension of polydisperse spheres, with $A(k)$ and the ISF now being suitably-weighted sums over the $M$ particles species $i= 1$ to $M$:
\begin{eqnarray}
A(k) & = & \sum_i^M A_i(k),  \label{eq:Aik}\\
f(k,\tau) & = & \sum_i^M C_i(k)f_i(k,\tau) \;\;\mbox{with} \label{eq:fik}\\
C_i(k) & = & \frac{A_i(k)}{A(k)},\;\;\mbox{and} \label{eq:Cik}\\
f_i(k,\tau) & = & \exp{[-D_i k^2\tau]} \;\;\mbox{with} \;\;D_i = \frac{k_B T}{6\pi\eta R_i}. \label{eq:Dik}
\end{eqnarray}

To interpret results obtained by fitting these expressions to data, we need to understand how the population weights, $\{C_i(k)\}$ in Eq.~\ref{eq:fik}, scale with particle radius, $R$. Interpretation of image intensity fluctuations from a coherent (heterodyne) scattering perspective implicitly predicts\cite{Multiscale_Review} a scaling of $A(k) \propto R^6$ for Rayleigh scatterers,\footnote[3]{What is clear from the heterodyne DDM literature is that the image intensity scales as $NR^3$; but additional steps are needed to show that $A(k) \sim NR^6$,} but it is unclear how the use of incoherent illumination affects this scaling.

To analyse size scaling in the incoherent limit, note that $a(k)$ in Eq.~\ref{eq:Ak} is the two-dimensional (2D) Fourier Transform (FT) of $a(r)$, the intensity pattern of the image of one particle centred at the origin of the image plane (with radial coordinate $r$ only in the case of circular symmetry).\cite{aniso_DDM} For a homogeneously fluorescent particle that is much smaller than the depth of focus, $a(r)$ should, to a good first approximation, be given by the 2D projection of a solid sphere (mathematically, a 3-ball, $B_3$) onto the equatorial plane, ${\cal P}_2(B_3)$, transmitted through the microscope's optics. The Projection-Slice Theorem states that the 2D FT of a projection of a 3D object is given by a slice (perpendicular to the projection) through the origin of the FT of the 3D object.\cite{Janaswamy2020} So, the FT of ${\cal P}_2(B_3)$ is $\frac{4}{3}\pi R^3 P(kR)$ with the $P(kR)$ in Eq.~\ref{eq:formfac}, only now $k$ is the magnitude of wave vectors in a 2D rather than 3D Fourier space. In the dilute limit, where $S(k) \to 1$, Eq.~\ref{eq:DICF} becomes  
\begin{equation}
\label{eq:DICF2}
    g(k, \tau) = \underbrace{2N \rho^2 \left[ \frac{4}{3}\pi R^3 P(kR) \right]^2}_{A(k)} \left[1 - f(k, \tau)\right] + B(k),
\end{equation}
with contrast density $\rho$ (e.g., dye concentration in fluorescence), assumed here to be homogeneous and the same for all particles.

In phase contrast imaging, the image is a projection of the optical path length, so can again be approximated by ${\cal P}_2(B_3)$. The bright-field image in the geometric limit is a shadowgraph which can be approximated by $I_0 - \beta {\cal P}_2(B_3)$, where $I_0$ and $\beta$ are constants. In either case, Eq.~\ref{eq:DICF2} is recovered. Quite generally, whenever the contributions of a particle's volume elements to the image intensity contrast are additive, one finds 
\begin{equation}
    A(k) \sim NR^6 P^2(kR) \sim \phi R^3 P^2(kR). \label{eq:R3scaling0}
\end{equation}
Since $P(kR) \to 1$ at the low $k$ accessed in DDM, this predicts an $NR^6$ or $\phi R^3$ scaling of signal with particle size.

The above discussion readily generalises to arbitrary-shaped anisotropic particles when there are enough (independently oriented) particles to  sample the orientational distribution, or when their rotational diffusion is fast compared to the relevant timescales in a DDM experiment. In this limit, a slice through the spherically-symmetric orientationally-averaged 3D form factor is the single particle contribution to $a(k)$ in Eq.~\ref{eq:Ak}. The well-known Guinier approximation to the low $k$ form factor,\cite{Guinier1955} Fig.~\ref{fgr:form_factor}, then gives $a(k) \sim V_p^2 e^{-k^2R_g^2/3}$, where $R_g$ ($= \sqrt{3R^2/5}$ for a sphere) is the particle's gyration radius. Now, the DDM signal $A(k)$ scales as $NV_p^2 \sim \phi V_p$, which is the generalisation of Eq.~\ref{eq:R3scaling0}. 

Eq.~\ref{eq:DICF2} does not take into account the finite depth of field in the $z$ direction. To do so, note first that for bright-field imaging at low numerical apertures typical in DDM, $k_z \lesssim \mathrm{NA} \times k$, so that the longitudinal dynamics are much slower than the in-plane dynamics. We can then take $f(k, k_z, \tau) \approx f(k, 0, \tau)$.\cite{Giavazzi2009DDM,giavazzi_digital_2014} The effect of a finite depth of field on $A(k)$ and limited lateral resolution can be included by convolving the real-space density with the optical point-spread function, or multiplying the density by the optical transfer function, OTF($k, k_z$), in reciprocal space to obtain 
\begin{equation}
\label{eq:DICF_OTF}
    A(k) = 2N \rho^2 \left[ \frac{4}{3}\pi R^3 \right]^2 \underbrace{\int |\mathrm{OTF}(k, k_z)|^2 \; P^2\left(\sqrt{k^2 + k_z^2} \;R\right) dk_z}_{P^2_\mathrm{eff}(k, R)}.
\end{equation}
The averaging of $P^2(k,k_z)$ over $k_z$ weighted by $|\mathrm{OTF}(k, k_z)|^2$ gives a squared effective form factor, $P^2_\mathrm{eff}(k, R)$, so that
\begin{equation}
    A(k) \sim NR^6 P^2_\mathrm{eff}(k, R) \sim \phi R^3 P^2_\mathrm{eff}(k, R), \label{eq:R3scaling}
\end{equation}
preserving the $R^6$ scaling. Substitution into Eq.~\ref{eq:fik} gives
\begin{equation}
    f(k,\tau) \sim \sum_i^M \phi_i R_i^3 P^2_\mathrm{eff}(k, R_i) f_i(k,\tau), \label{eq:fweighting} 
\end{equation}
where $\phi_i$ is the volume fraction of species $i$. 
If all species are small enough such that $k R_i \ll 1$, then $P \to 1$ for all species and $P_\mathrm{eff}(k, R_i) = P_{\rm eff}(k)$ becomes the square of the projection of OTF onto the $k_z = 0$ plane, dropping out of $f(k, \tau)$ so there will be no form-factor minima effects. However, for larger particles the form factor $P(kR_i)$ can drop noticeably below unity over the range of $k$ probed by DDM, with a corresponding $k$ dependence and overall magnitude drop in $P_\mathrm{eff}(k, R)$.
Furthermore, our treatment neglects refraction at the particle-liquid interface, which for large particles can significantly modify their contrast in bright-field imaging. So, $R^6$ scaling likely fails for large particles.

\section{Materials and Methods}

\subsection{Experimental}
\label{ssec:Materials}

We used polystyrene spheres, which have been routinely characterised using DDM.\cite{CerbinoDDMReview, Giavazzi2009DDM, Cerbino2008DDM} Dispersions from Thermo Scientific 5000 series with sizes (diameters here and throughout) of \SI{60}{\nano\meter}, \SI{120}{\nano\meter}, \SI{240}{\nano\meter}, \SI{500}{\nano\meter}, \SI{1.1}{\micro\meter} and \SI{2.1}{\micro\meter} were diluted using Milli-Q water to give stock solutions of various concentrations, from which we prepared various bimodal or trimodal mixtures. Note that only the smallest of these particles are Rayleigh scatterers. Samples were loaded into 0.4\x{}4\x{}50~mm glass capillaries (Vitrocom Inc.) and sealed with Vaseline to prevent evaporation. Bright-field videos were captured using a Nikon Ti-E inverted microscope with a Hamamatsu Orca Flash 4.0 camera. We imaged far from the sides of the capillary and \SI{100}{\micro\meter} from the base. For each measurement a series of five videos were captured immediately after loading to minimise sedimentation. Each video is 5000-6000 frames of 512\x{}512 pixels. Specific choices of frame rate and objective, detailed below, reflect these considerations:
\begin{itemize}
    \item Pixel size -- DDM does not require resolvable particles and pixel $\gtrsim$ particle size typically gives the best results: large pixels mean lower $k$, minimising form factor effects.
    \item Frame rate -- Chosen to capture $\lesssim4$~Gb 16-bit TIFF data including both short- and long-time plateaus of the ISF. In this work the range is 100-400~fps.
\end{itemize}
Small changes to the settings did not in general significantly impact results except in the extreme cases treated in Section~\ref{sec:Trimodal}. 

The DICF is extracted from videos using previously-described LabView software.\cite{Jochen_Citation} The uncertainty in the DICF is estimated as the standard error on the mean from the azimuthal averaging of $k$. A more theoretical approach requires quantifying the variance of background image intensity;\cite{DDM_Uncertainty} but such rigour is not needed here and is likely too demanding for general application.  

\subsection{Data fitting}
The extracted DICFs are fitted to Eq.~\ref{eq:DICF} with model $f(k, \tau)$ to extract the diffusivities, $\{D_i\}$; here we outline the case of bidispersity. To decide a suitable range of $k$ for analysing each system, we carried out DDM experiments with the two individual populations of particles used to make a bidisperse sample, and used independent 1D fits to each $k$ dataset to extract the $k$-dependent average diffusivities $D_1(k)$ and $D_2(k)$. The range of $k$ values over which these are both flat to within noise is used for all subsequent data fitting with these particles and microscope settings.

\subsubsection{Least Squares} \label{sec:leastsqintro}
Global least-squares (LS) fits at all $k$ within the chosen range are performed simultaneously using the Levenberg-Marquadt algorithm implemented in Scipy.\cite{scipy} Other algorithms often failed to converge or returned biased diffusion coefficients in multimodal fits. We fitted $g(k,\tau)$ with $\{A(k)\}$ and $\{B(k)\}$ as free parameters for each $k$ and $k$-independent fit parameters in $f(k, \tau)$ (e.g.~diffusivities). Three models of the ISF were used:
\begin{enumerate}
    \item $f(k, \tau) = \exp(-Dk^2 \tau)$ -- for monodisperse diffusing particles. 
    \item $\ln(f(k, \tau)) = -\mu_1 k^2 \tau + \mu_2 (k^2 \tau)^2 /2 - \mu_3 (k^2 \tau)^3 / 6 + ...$ -- the cumulant expansion\cite{Frisken2001Cumulant, 2015Cumulant, FinsySizingReview} typically used to extract the mean diffusivity ($\mu_1$) and polydispersity (from $\mu_2$) in low-polydispersity monomodal samples. 
    \item $f(k, \tau) = C_1 \exp(-D_1 k^2 \tau) + (1 - C_1)\exp(-D_2 k^2 \tau) $ -- for two monodisperse populations with diffusivities $D_1$ and $D_2$ contributing fractions $C_1$ and $1 - C_1$ to the signal respectively. 
\end{enumerate}
Figure~\ref{fgr:simulated}a illustrates the information extracted by fitting these models to a simulated ISF from a bidisperse distribution of diffusivities. Model (1) finds an essentially meaningless `average' that misses both populations. Model (2) suffers from the same problem as far as the mean value is concerned, but gives a credible description of the notional `polydispersity'. Model (3) returns more or less correct average sizes and contributions for the two populations, but does not deal with the polydispersity within each.

\subsubsection{CONTIN}
\begin{figure}[t]
\centering
  \includegraphics[width=\linewidth]{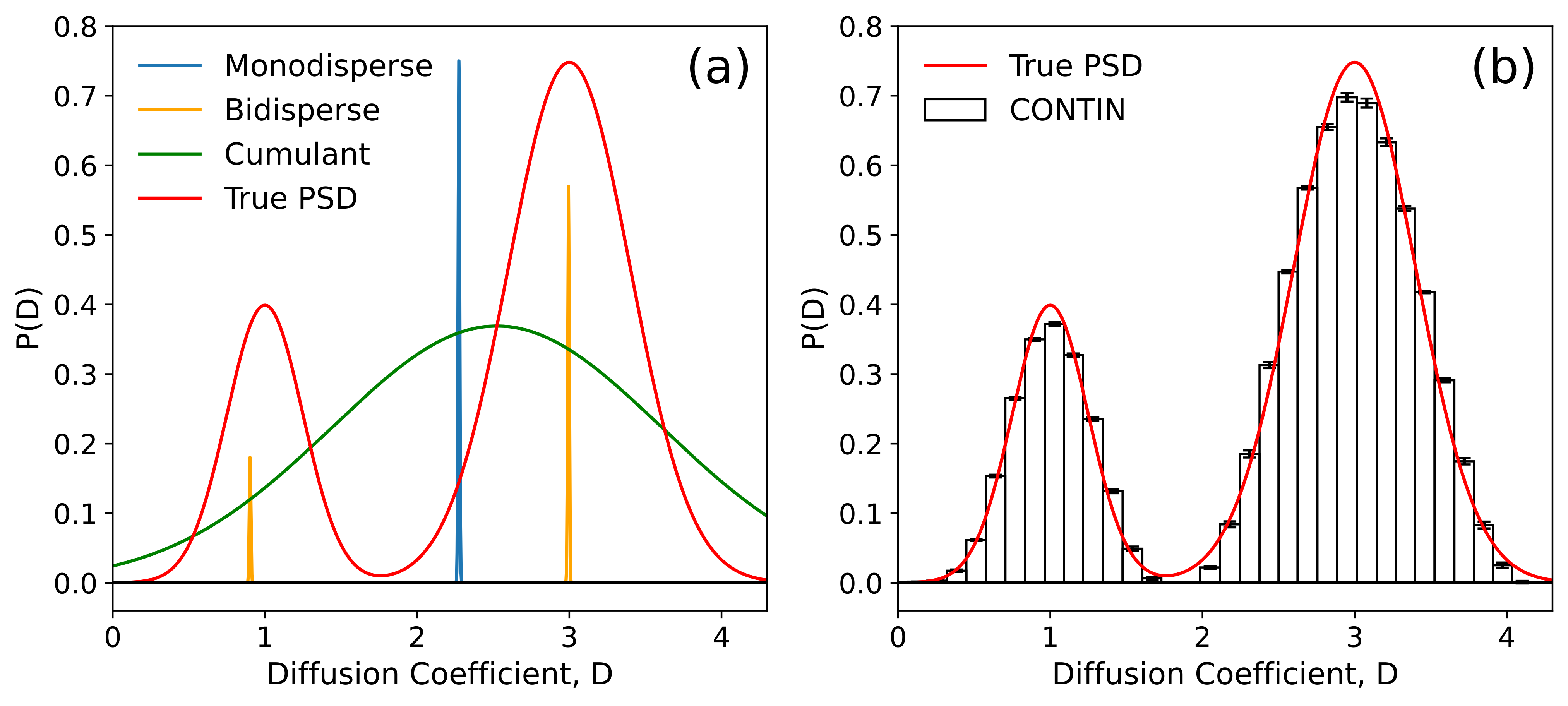}
  \caption{Extracted particle diffusivity distributions, $P(D)$, from a simulated ISF based on a defined bimodal $P(D)$ (red curves, $\mu_1 = 1$, $\mu_2 = 3$, $\sigma_1 = 0.25$, $\sigma_2 = 0.4$, $C_1 = 0.25$). ISF generated with Gaussian noise at each $f(k, \tau)$; $\sigma=10^{-5}$. (a) Graphical representation of output from three different least squares fits; monodisperse (blue), bidisperse (orange), and a cumulant expansion (green). (b) Output from a CONTIN fit.}
  \label{fgr:simulated}
\end{figure}

The CONTIN algorithm\cite{Provencher1982AEquations, Provencher1982Program} has long been used to extract the distribution of diffusivities from measured ISFs in DLS. It returns the size of the contributions to the composite ISF, $\{C_i(k)\}$ in Eq.~\ref{eq:fik}, for a finite number of user-determined bins, giving a normalised distribution of diffusivities weighted according to eq.~\ref{eq:R3scaling}, the `Particle Diffusivity Distribution (PDD)', $P(D)$, which is linked to the PSD by the Stokes-Einstein relation. The presence of noise in the data renders this inverse problem ill-posed. CONTIN deals with this by `regularisation',\cite{Twomey1977} i.e., balancing fit quality against parsimony by favouring a certain degree of `smoothness' in $P(D)$. We investigated a variety of criteria for optimising this balance (via tuning $\alpha$, the `regularisation parameter' in CONTIN), including the L-curve\cite{CONTIN_LCurve} and reduced-$\chi^2$ statistic. However, Provencher's method of selecting $\alpha$ by comparing the impact of regularisation and the noise in the data, which is implemented as part of CONTIN,\cite{Provencher1982AEquations,Provencher1982Program} was consistently found to work best. 

Figure~\ref{fgr:simulated}b illustrates the result of this procedure in fitting the ISF from a bimodal distribution of diffusivities, inputting only the desired binning of the output histogram. CONTIN's regularisation selection works exceptionally well because the only noise applied to the simulated $f(k, \tau)$ is Gaussian with a known amplitude ($\sigma = 10^{-5}$), and is independent for each $k$-$\tau$ pair. With real experimental noise, the selection of $\alpha$ is generally more difficult.

CONTIN is designed for linear problems, but extracting $\{C_i\}$ from the DICF by fitting Eqs.~\ref{eq:Aik}-\ref{eq:Dik} to Eq.~\ref{eq:DICF} is non-linear, because $A(k)$ and $B(k)$ are unknown. We therefore must estimate these parameters before using CONTIN. One approach is to perform a least-squares fit of $g(k, \tau)$ with an approximate model (e.g.~a cumulant expansion) and use the returned $A(k)$ and $B(k)$ to extract an ISF to pass on to CONTIN. However, we found that this encoded the approximate model into the CONTIN results. 

Alternatively, since $g(k, \tau \rightarrow 0) = B(k)$ and $g(k, \tau \rightarrow \infty) = A(k) + B(k)$, the long- and short-time `plateau values' of $g(k,\tau)$ can in principle give $A(k)$ and $B(k)$.\cite{CerbinoDDMReview} Under practical experimental conditions, however, it is often challenging to access one or the other of these limits. For us, the long-time plateau $g(k, \tau \rightarrow \infty)$ is typically accessible whilst the  short-time plateau $g(k, \tau \rightarrow 0)$ is difficult to reach even at the highest frame rates, and errors in estimating $B(k)$ can significantly impact results.\cite{DDM_Uncertainty} 

Instead, we extract $B(k)$ by fitting the first 10-15 time points of $g(k, \tau)$ to a second order polynomial of the form 
\begin{equation}\label{eq:Taylor_Expansion}
    g(k, \tau) \approx B(k) + \beta_1(k) \tau - \beta_2(k) \tau ^2
\end{equation}
for each $k$. This form can be justified by substituting $f(k, \tau) = \int P(D) \exp(-D k^2 \tau) \; {\rm d}D$, the continuum version of Eq.~\ref{eq:fik}, into Eq.~\ref{eq:DICF} and Taylor expanding around $\tau = 0$, using that $P(D)$ is normalised. For completeness, we find $\beta_1(k) = k^2 A(k)\int P(D) D \; dD$ and $\beta_2(k) = \frac{1}{2} k^4 A(k) \int P(D) D^2 \; dD$. The fitted value of $B(k)$ can then be subtracted from the average of the final 10-15 data points to obtain $A(k)$. With this, the ISF can be extracted from $g(k, \tau)$ and passed to CONTIN with an uncertainty estimate based on propagation of errors in $g(k, \tau)$, the standard error of the data points averaged for $A(k)$, and the polynomial fit uncertainties for $B(k)$.

\section{Results: scaling of DDM Signal with particle size} \label{sec:Scaling}
To verify Eq.~\ref{eq:R3scaling}, we performed DDM experiments on quasi-monodisperse suspensions with a range of radii, $R$. A sample of each suspension from Section \ref{ssec:Materials} was diluted to a mass fraction $\psi = 10^{-5}$. Five bright-field videos of each were captured at 200~fps using a 20\x/0.5 objective without binning, giving \SI{325}{\nano\meter} pixels. Using phase-contrast illumination produced equivalent results. A least squares 3rd order cumulant fit of the DICF from each video gives $A(k)$, $B(k)$, and average diffusion coefficient. Identical microscope settings ensured that changes in $A(k)$ are solely due to particle size, and there is no measurable systematic trend in average intensity with $R$ so turbidity is negligible in all cases. Each $A(k)$ was normalised by that of the \SI{60}{\nano\meter} particles, for which $P(kR) \approx 1$ for all $k$. This removes the significant $k$ dependence of the OTF. The range $ \SI{1.0}{\per\micro\meter} \leq k \leq \SI{2.5}{\per\micro\meter}$ was used for all videos to remove the effect of any additional $k$ dependence. To isolate the power-law dependence on particle size, we remove the the form factor contribution to $A(k)$ by dividing by the squared form factor for a sphere, $P^2(kR)$ (Eq.~\ref{eq:formfac}). 

\begin{figure}[t]
 \centering
 \includegraphics[width=\linewidth]{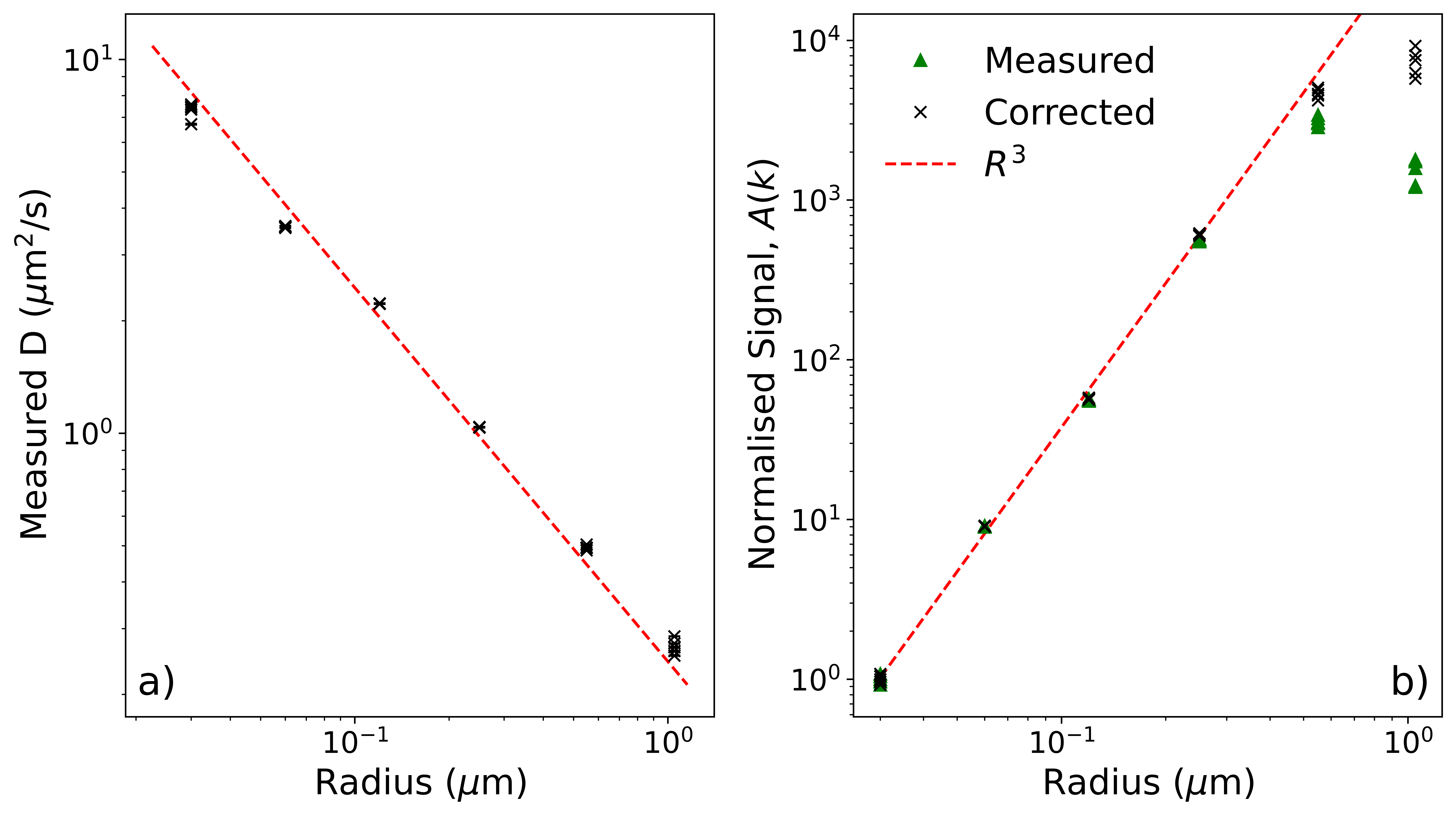}
 \caption{DDM results for polystyrene spheres of different sizes. (a) Extracted diffusion coefficients as a function of manufacturer provided radius. Dashed line shows Stokes-Einstein prediction (b) Average normalised (see main text) $A(k)$ against particle radius. Green triangles indicate average $A(k)$ as extracted from videos. Black crosses are the same data corrected for form factor effects. The dashed line has slope 3. In both plots there are 5 points for each size, which often overlap.}
 \label{fgr:scalingBF}
\end{figure}

For all sizes, the measured diffusivity agrees with the Stokes-Einstein value, Fig.~\ref{fgr:scalingBF}a. We can therefore size particles over two orders of magnitude of $R$ without changing experimental settings, even though $A(k) < B(k)$ for the smaller particles.

Figure \ref{fgr:scalingBF}b shows that for $R \lesssim \SI{0.55}{\micro\meter}$, $A(k) \propto R^3$ at constant $\psi$ (and therefore $\phi$), verifying Eq.~\ref{eq:R3scaling}. Since $N \propto \psi/R^3$ and we have previously confirmed experimentally\cite{arlt_painting_2018} that $A(k) \propto N$, this is equivalent to a signal of $R^6$ per particle, matching DLS scaling (assuming Rayleigh scatterers) and our theoretical prediction. This $R^6$ scaling sets the weight of signal contributions in the case of mixtures with multiple particle sizes. Note however, that the (temporal) intensity fluctuations which form the basis for DLS measurements also scale as $R^6$, whereas the (spatio-temporal) fluctuations that form the basis of the DDM measurements only scale as $R^3$. So, it is experimentally much less challenging to capture intensity fluctuations faithfully for DDM. 

For the largest particles, $A(k)$ increases with $R$ slower than $R^3$, even after correcting for form factor effect, Fig.~\ref{fgr:scalingBF}b. As already suggested at the end of Section \ref{sec:PolydisperseDDM}, this is not unexpected as several assumptions of our simple model start to fail.

\section{Results: bidisperse systems} \label{sec:Bidisperse}
The simplest multimodal suspension is bimodal, with milk\cite{MilkBimodal} being an everyday example. We mixed $10^{-4}$ mass fraction dispersions of \SI{240}{\nano\meter} and \SI{1.1}{\micro\meter} particles (size ratio $\approx$ 1:4.6) to produce bidisperse mixtures in which the small particles should contribute between 1\% and 99\% of the signal to the ISF according to Eq.~\ref{eq:R3scaling}; Table~\ref{tbl:bidisperse_mix}. Videos of each sample and of the parent populations were captured at 100~fps, with a 10\x/0.3 objective and 1.5\x{} extra magnification (pixel size \SI{433}{\nano\meter}). Examples of the extracted DICFs and fits are shown in Appendix~\ref{appendix:DICF_examples}.

\begin{table}[b]
\small
 \caption{ \SI{240}{\nano\meter} and \SI{1.1}{\micro\meter} particle mixtures used in section~\ref{sec:Bidisperse}.}
 \label{tbl:bidisperse_mix}
 \begin{tabular*}{0.48\textwidth}{@{\extracolsep{\fill}}ll}
    \hline
    Large-particle mass fraction & Expected small-particle ISF contribution \\
    \hline
    70\% & 1\% \\
    50\% & 2\% \\
    25\% & 5\% \\
    10\% & 13\% \\
    5\% & 25\% \\
    2.5\% & 40\% \\
    1\% & 63\% \\
    0.5\% & 77\% \\
    0.1\% & 95\% \\
    0.02\% & 99\% \\
    \hline
 \end{tabular*}
\end{table}

\subsection{Least Squares Fits}
\begin{figure}[t]
\centering
  \includegraphics[width=\linewidth]{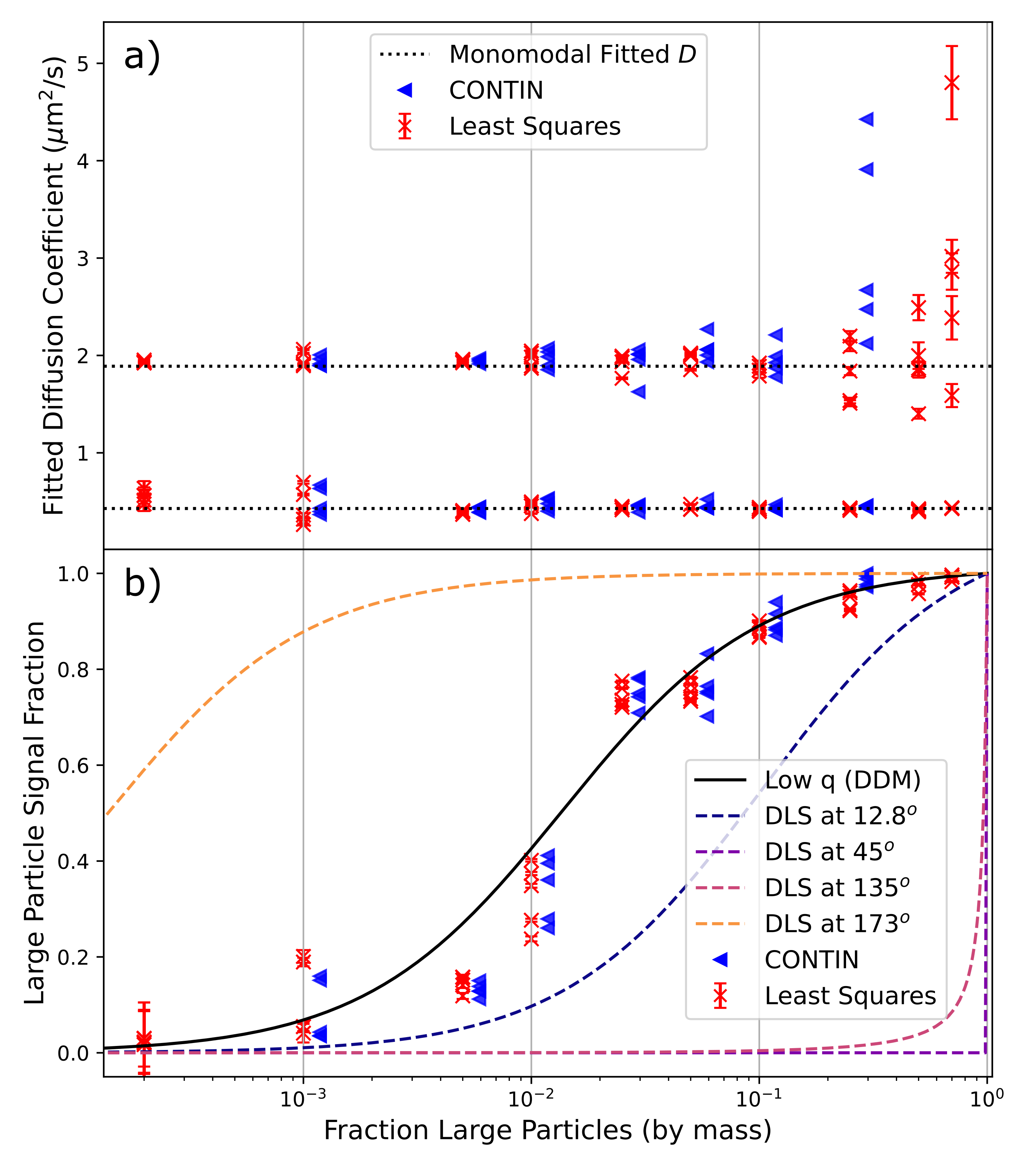}
  \caption{Results of DDM analysis of 240~nm/1.1\micron{} sphere mixtures with different compositions, showing five measurements at each composition. Red crosses indicate results of least-squares fits to an explicit bimodal PDD, blue triangles are extracted from CONTIN fits. CONTIN results are shifted slightly along the x-axis for clarity, and are only plotted where each population is expected to contribute $\gtrsim 5\%$ of the DICF. a) Diffusion coefficients (points) compared to average values for the monomodal suspensions (dotted lines). b) Signal fraction from large particles (points), and theoretical expectations for DDM, Eq.~\ref{eq:R3scaling} (solid line) and DLS at various angles (dashed lines).}
  \label{fgr:Bidisperse_LS_fig}
\end{figure}

Fitting model 3 in Section~\ref{sec:leastsqintro} to our data, which assumes bidispersity, we extract the mean diffusion coefficient and the relative contribution of each population to the ISF, Fig.~\ref{fgr:Bidisperse_LS_fig}. Comparison with values obtained from fitting a monodisperse model to the unmixed samples, Fig.~\ref{fgr:Bidisperse_LS_fig}a, shows that the method works well provided that the `low-signal component'\footnote[4]{I.e., the component of the bidisperse suspension that contributes $< 50\%$ of the $g(k, \tau)$ signal; because of the $NR^6$ scaling, this is {\it not} the `minority component' by number.} contributes at least $\approx 2\%$. We found little quantitative difference in taking $C_1$ in model 3 to be constant or allowing it to vary with $k$, confirming minimal form factor effects. Practically, allowing $C_1$ to vary increased processing time and occasionally caused issues with convergence.

The spread of the five measurements of each sample shows that fit uncertainties shown by the error bars are underestimated, although this is not indicated by fit statistics (reduced-$\chi^2 \approx 1$). The uncertainties are comparable across different fitting algorithms, including Minuit’s MINOS error estimation,\cite{Minuit} which accounts for correlations between fit parameters. The underestimate could be due to correlations in $g(\vec{k}, \tau)$ that are not accounted for when calculating DICF uncertainty or in the fitting algorithms. We did not investigate these uncertainties because the final error bar in the fitted diffusivities (and therefore size) is clearly defined by variability between measurements.  

Note that using DLS to obtain the correct diffusivities for the two populations would only be possible if the scattering angle, $\theta$, is optimised to avoid the form factor minima of each. To highlight this, we plot the theoretical fractional contribution of large particles to the DLS signal at different $\theta$ for DLS using a \SI{532}{\nano\meter} laser and polystyrene spheres (refractive index = 1.59), Fig.~\ref{fgr:Bidisperse_LS_fig}b, with $\theta = \SI{12.8}{\degree}$ and \SI{173}{\degree} being typical of some popular commercial devices. Note that these curves assume that each population is monodisperse; any polydispersity would cause significant shifts.

By contrast, there is a unique theoretical prediction for DDM, which is calculated only using Eq.~\ref{eq:R3scaling} from the mean sizes of the two populations, with the latter being extracted from the same experiment, Fig.~\ref{fgr:Bidisperse_LS_fig}. Since we remain at low $k$ far from the form factor minimum, polydispersity of the individual populations can be neglected in calculating this curve. Thus, our protocol can give direct information on the composition of the sample. 

If the size ratio of our bidisperse suspension is reduced from 1:4.6 to 1:2, fitting DDM data to model 3 yields a significantly biased diffusivity for the smaller particles even when they contribute as much as 40\% of the signal (Appendix~\ref{subapp:Factor_2_res}). At this size ratio, the corresponding timescales for decorrelation are too close for them to be separated cleanly. There is some indication of local minima in the $\chi^2$ minimisation, suggesting that alternative approaches may improve results. However, we also observed this bias in analysis of simulated bimodal ISFs when the peaks in $P(D)$ begin to overlap, so this may represent a more general limitation that is not unique to DDM. DDM therefore cannot be solely relied upon to size bidisperse samples with such low size ratios.

\subsection{CONTIN Analysis}
Least-squares fitting delivered the correct mean diffusivities of the two populations and their number ratio by assuming bidispersity. Alternatively, the cumulant model (model 2 in Section~\ref{sec:leastsqintro}) can be fitted to the data without this assumption to obtain a single mean and polydispersity (see Appendix~\ref{appendix:Cumulant_Fits}), with no indication of bidispersity or poor fit quality. To do better, we turn to CONTIN.

\begin{figure}[t]
\centering
  \includegraphics[width=0.95\linewidth]{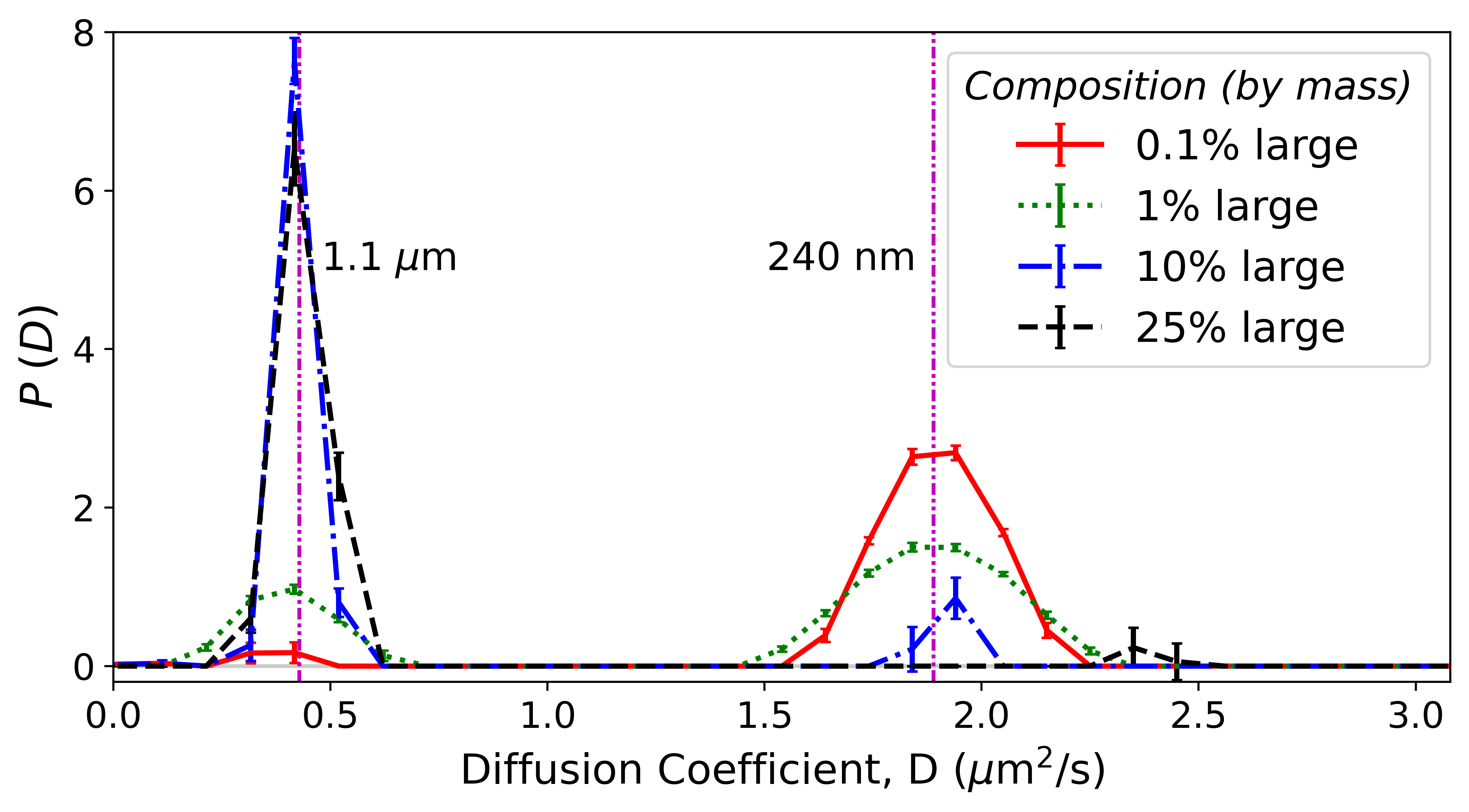}
  \caption{CONTIN results for various 240~nm/1.1\micron{} sphere mixtures (see legend) with expected signal contributions from large particles of 5\%, 37\%, 87\%, and 95\% respectively. Purple vertical lines show average diffusivities from least-squares fits to the monodisperse suspensions.}

  \label{fgr:CONTIN_Bidisperse}
\end{figure}

CONTIN delivers $P(D)$, the particle diffusivity distribution (PDD) histogram on a predefined grid of 60 linearly spaced bins in the interval $\SI{0.01}{\micro\meter^2\second^{-1}} \leq D \leq \SI{5}{\micro\meter^2\second^{-1}}$. Figure~\ref{fgr:CONTIN_Bidisperse_Full}  (Appendix~\ref{subapp:Factor_4_CONTIN}) shows the result for each sample in Table~\ref{tbl:bidisperse_mix}. Figure~\ref{fgr:CONTIN_Bidisperse} shows the PDDs from the third video of each mixture in which the large particles contribute 0.1\%, 1\%, 10\%, and 25\% of the particle mass (or 5\%, 37\%, 87\%, and 95\% of the signal).  

This analysis convincingly returns a bimodal distribution of diffusivities provided that the contribution of low-signal component to the DICF remains $\gtrsim 5\%$, comparable but slightly more stringent than for least-squares fitting. This is because too small a contribution to $f(k, \tau)$ from either species will be removed as `noise' by the CONTIN regularisation algorithm, whilst least squares will always return two sizes -- fitting noise if necessary.

Fitting the weighted sum of two Gaussian distributions to the returned PDDs for each video with $\geq 5\%$ minority signal yields the mean diffusivity and relative signal contribution of each population, Fig.~\ref{fgr:Bidisperse_LS_fig}. The variation in these properties is comparable to the equivalent least-squares values, and the more stringent signal contribution requirements are visible as the signal reaches $\approx 5\%$. These fits also return a polydispersity; but there are significant run-to-run variations in the fitted PDD at each composition, Fig.~\ref{fgr:CONTIN_Bidisperse_Full}, because the regularisation parameter $\alpha$ is highly noise sensitive. However, CONTIN fits the data to an integral of the PDD, so that there is {\it a priori} reason to surmise that the area of each peak may be far less noise sensitive than either the peak width or height. The peak area is a measure of the (weighted) number of particles, Eq.~\ref{eq:R3scaling}. Figure~\ref{fgr:Bidisperse_LS_fig}b validates this surmise. So, a CONTIN analysis is able to deliver the sizes and relative number of the two populations in our 1:4.6 bidisperse suspension.

We also tested CONTIN analysis for a bidisperse suspension in which the two populations differ in size by only a factor of 2. The method again returns a bimodal diffusivity distribution with essentially the same means as least-squares fits (including the aforementioned bias) whenever the low-signal component contributes $\gtrsim 5\%$ of the DICF, Fig.~\ref{fgr:CONTIN_factor_two} (Appendix~\ref{subapp:Factor_2_res}).

\section{Spatial aspects of DDM} \label{sec:Trimodal}

\begin{figure}[t]
\centering
  \includegraphics[width=0.8\linewidth]{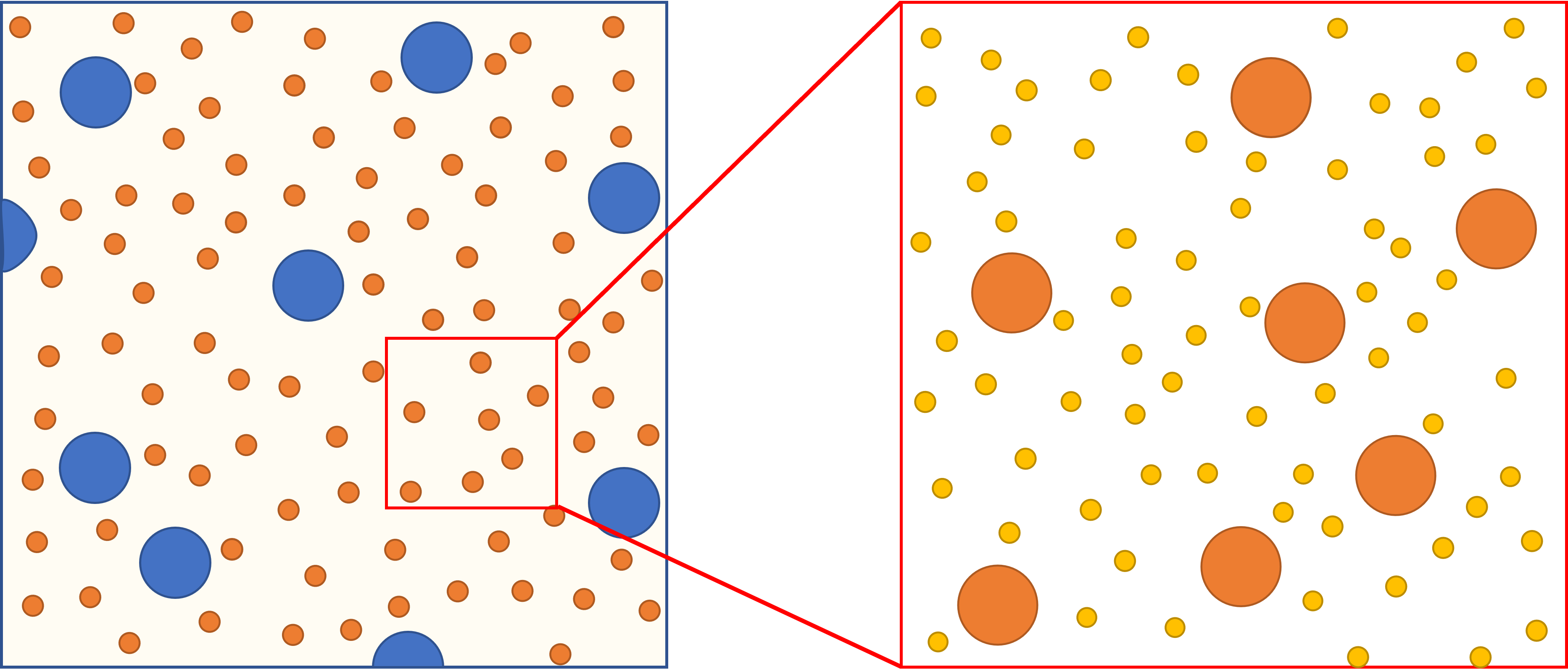}
  \caption{Schematic of a trimodal suspension showing how by selecting a suitable region of interest, we can enhance the DDM signal from smaller particles by removing the contribution from large particles.}
  \label{fgr:trimodal_illustration}
\end{figure}

The $NR^6$ scaling in DLS and DDM means that even a low concentration of large particles will dominate the signal and render it difficult, if not impossible, to detect smaller species. Thus, for example, in our 1:4.6 bidisperse suspension, we need at least 75\% mass fraction of the smaller species to contribute at least 5\% of the DICF, Table~\ref{tbl:bidisperse_mix}, for this population to show up in the PDD from a CONTIN analysis. Such considerations are important, e.g., when sizing biomedical nanoparticles, where buffers at physiological ionicity often lead to aggregation. The presence of micron-sized aggregates leads to highly distorted PSDs\cite{Hondow2012} or even irreproducible results when DLS was used to size nanoparticles.\cite{bionano2016} We next show how to use DDM to size one or more populations of small particles in the presence of a numerically-minor population of large particles that dominate the signal.

The key is to make use of the spatial information encoded in the images collected in a DDM experiment. The numerical minority of the largest particles means that they are relatively sparse in the images. So, it should be possible to analyse selectively only those portions of the collected images from which these particles are essentially absent, Fig.~\ref{fgr:trimodal_illustration}. This can be accomplished either by combining dilution and control of magnification or by using a spatially-resolved analysis. We demonstrate these two approaches using a trimodal stock, Table~\ref{tbl:trimodal_mix}. 

\begin{table}[b]
\small
  \caption{Trimodal system composition for Section \ref{sec:Trimodal}.}
  \label{tbl:trimodal_mix}
  \begin{tabular*}{0.48\textwidth}{@{\extracolsep{\fill}}llll}
    \hline
    Particle Diameter & 60 nm & 240 nm & 1.1\micron{}\\
    \hline
    Weight Fraction & $10^{-4}$ & $10^{-6}$ & $10^{-6}$\\
    Signal Contribution, $C_i$ & 2\% & 1\% & 97\%\\
    Number Density (per mm$^3$) & $8\times 10^8$ & $1 \times 10^5$ & $2\times 10^3$\\
    \hline
  \end{tabular*}
\end{table}

First, we measured a sequence of samples obtained by successively diluting the stock by a factor of 3. The idea is to identify, if possible, a window of concentration in which the field of view typically does not include any of the largest particles, but the signal level from the smaller populations is still measurable. 

\begin{figure}[t]
\centering
  \includegraphics[width=\linewidth]{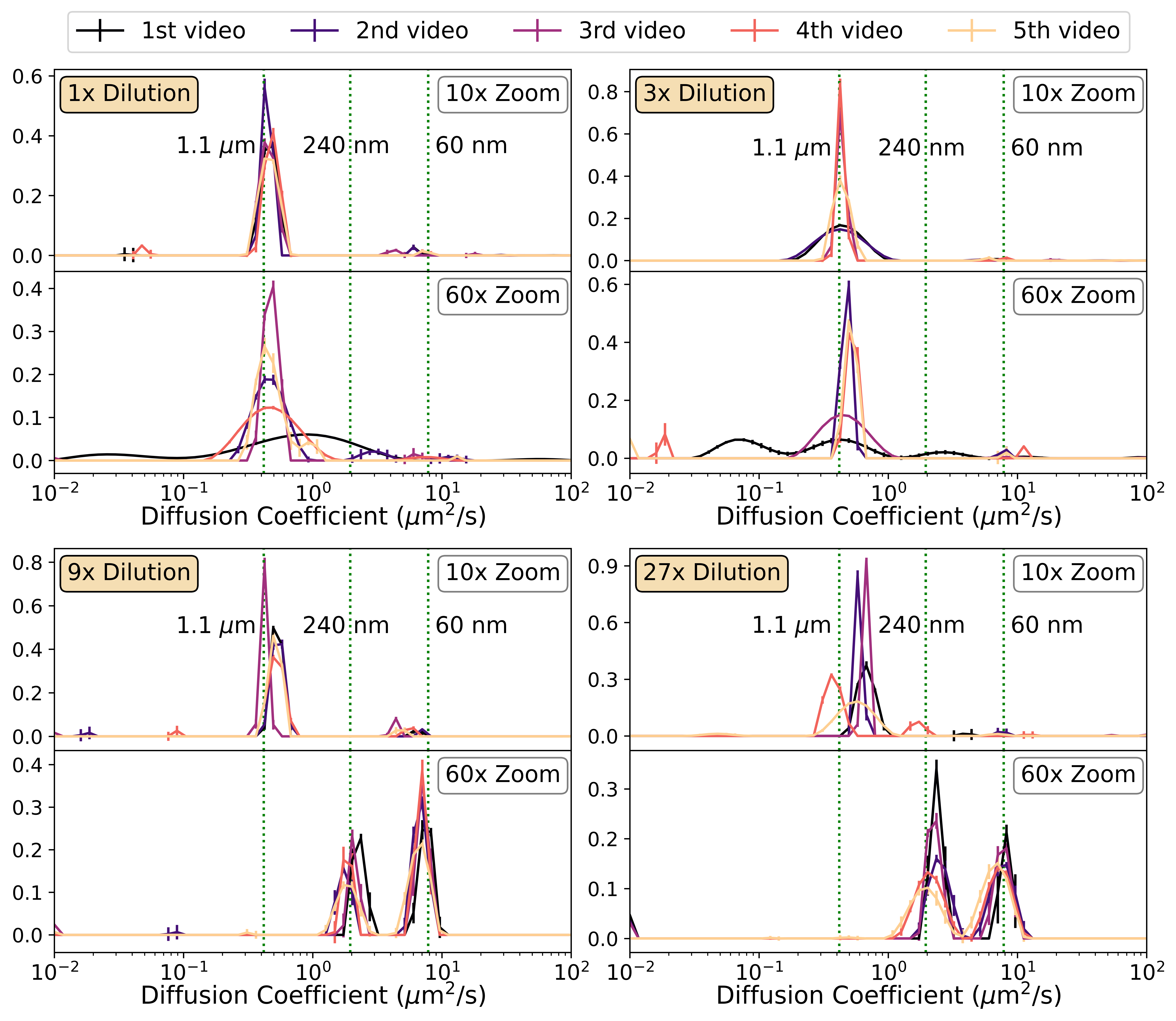}
  \caption{CONTIN fits to videos of the trimodal system defined in Table \ref{tbl:trimodal_mix} and dilutions. Each pair of graphs shows the result with 10\x{} magnification (top) and 60\x{} magnification (bottom) for the labelled dilution. Vertical dotted lines show expected peak positions.}
  \label{fgr:CONTIN_Trimodal}
\end{figure}

Videos of each dilution were captured at 400 fps using both 10\x/0.3 and 60\x/0.7 objectives without pixel binning, for pixel sizes of 650~nm and 108~nm respectively. For 512\x{}512 pixel images (the maximum square images possible at this frame rate with our camera) this corresponds to 2D imaging areas of $1 \times 10^5$\micron{}$^2$ and $3 \times 10^3$\micron{}$^2$ respectively. The exact volume imaged is difficult to estimate since the depth of field is strongly $k$ dependent;\cite{Giavazzi2009DDM, DDMTurbid} an estimate based on geometric optics\cite{Geometric_Optics} would be 9\micron{} and 1\micron{} for the 10\x{} and 60\x{} objective respectively.

By design, the number of modes found by the analysis should vary as dilution progresses, with the largest population disappearing to reveal smaller particles. With no {\it a priori} fixed number of modes to input to least-squares fitting, we used CONTIN, with the input being a grid of 60 logarithmically spaced bins for $\SI{e-2}{\micro\meter^2\second^{-1}} \leq D \leq \SI{e2}{\micro\meter^2\second^{-1}}$. To avoid logarithmically scaled bin heights, the quadrature weight of each bin is set to 1, so that the sum of bin heights describes the contribution from each population rather than the bin area.\cite{Provencher1982AEquations,Provencher1982Program} Figure~\ref{fgr:CONTIN_Trimodal} shows the fitted $P(D)$ from each repeat using the two different objectives with the stock suspension and three successive dilutions. 

With the larger field of view (10\x{} magnification) the large particles dominate the signal at all dilutions. At the higher (60\x{}) magnification, there is still no convincing evidence of the two smaller populations until we reach $3^2$-fold dilution, and their signals remaining robust at $3^3$-fold dilution. Again, there is significant variability in peak shape from run to run, but the overall picture is clear. Further dilution reduces the signal to the extent that peaks appear and disappear in the 5 repeats. 

\begin{figure}[t]
\centering
  \includegraphics[width=0.95\linewidth]{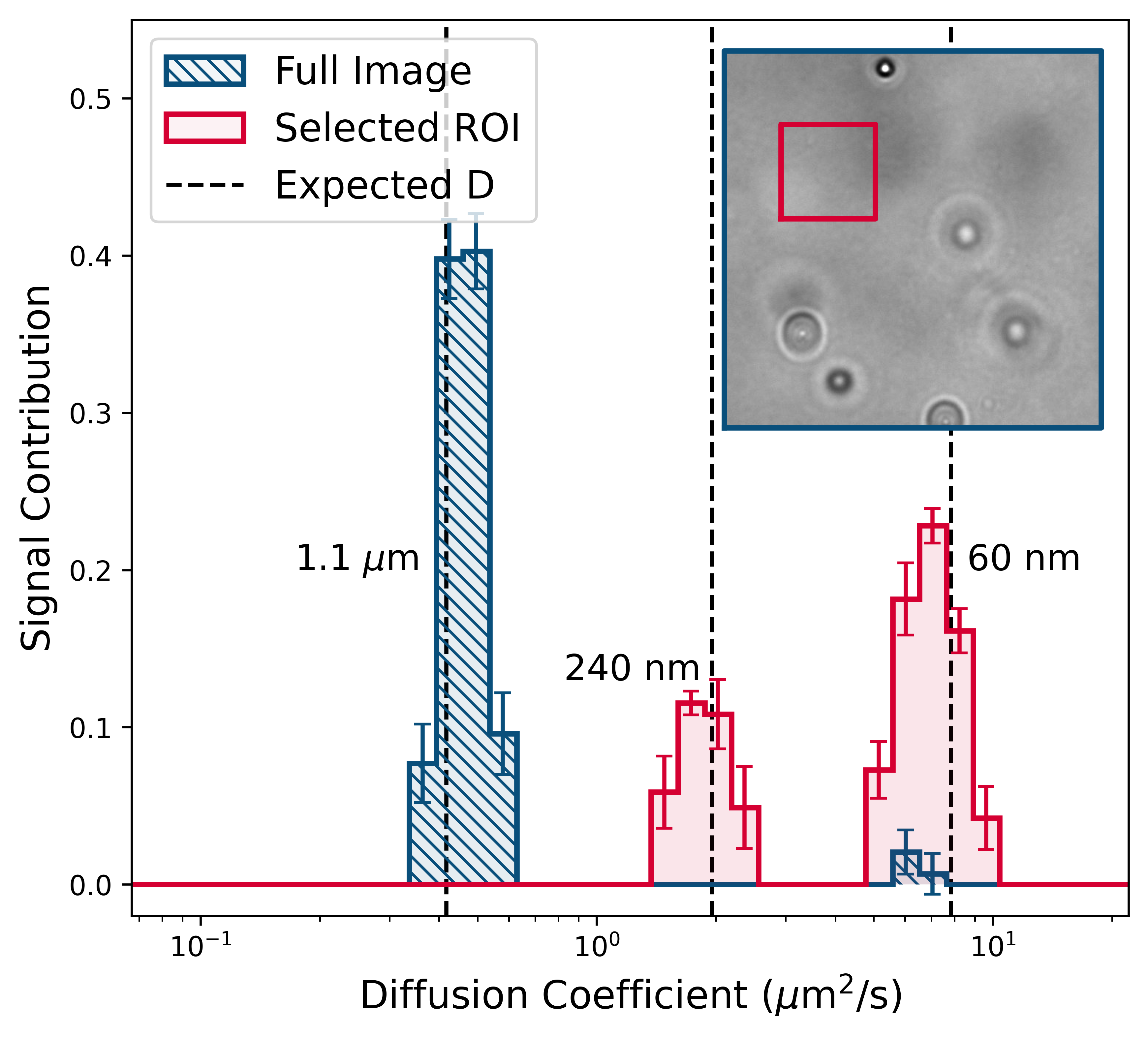}
  \caption{CONTIN fits to data extracted from 60x magnification videos of the (undiluted) mixture in table 2, and from a 128\x{}128 pixel selected ROI. Dashed vertical lines show the diffusivities of the monomodal suspensions. Inset shows example frame and the selected ROI.}
  \label{fgr:Trimodal_ROI}
\end{figure}

A disadvantage of the dilution method is that the optimal concentration window is rather narrow, and so can be easily missed in a real-life application. More robustly, one may eliminate the contribution from the largest particles by selecting an appropriate region of interest (ROI) for analysis from the original video. Figure~\ref{fgr:Trimodal_ROI} compares the PDDs obtained from one of the 60\x{} magnification videos of the stock suspension when we analysed the full video (512\x{}512 pixels), and when we analysed a smaller ROI (128\x{}128 pixels) chosen to exclude all large particles. Not surprisingly, the former PDD shows only the largest particles, while the latter shows the two smaller populations. 

Obviously, success depends on selecting and optimising an ROI. Figure~\ref{fgr:ROI_Full} in Appendix~\ref{subapp:ROI_Full} reports the PDD obtained as the size of the ROI is progressively reduced for 60\x{} magnification videos of the stock solution, again showing five runs at each stage. That the two smaller populations show up strongly after reducing the ROI by a factor of $4^2$ is consistent with diluting the stock by $3^2$-$3^3$ times to give optimal performance, Fig.~\ref{fgr:CONTIN_Trimodal}. The full data set, Fig.~\ref{fgr:ROI_Full}, illustrates the superiority of this method compared to dilution. Here, the user varies the ROI size and position in real time while analysing a single data set until correct sizing is achieved; dilution requires multiple experiments in which the user must `hit' the right dilution window and sample position by chance. 

As we already noted, the peak areas in the CONTIN output contain compositional information via the relative values of $A(k)$. Following the procedure described in Appendix \ref{appendix:ROI_composition}, we extracted relative volume fractions, finding $98.1 \pm 0.1$\%, $1.0 \pm 0.1$\%, and $0.9 \pm 0.1$\% for the \SI{60}{\nano\meter}, \SI{240}{\nano\meter}, \SI{1.1}{\micro\meter} populations, in excellent agreement with the known $10^{-4}:10^{-6}:10^{-6}$, Table~\ref{tbl:trimodal_mix}. 

Conceptually, our technique is similar to centrifuging out large particles prior to DLS. However, our methods need lower quantities of suspension and do not require physical processing, which could be relevant for sparse or delicate samples. Note also that our ROI selection may be compared with the use of spatially-resolved DDM to verify a theorem in active matter physics.\cite{Arlt2019}

\section{Summary and conclusions}

Our results show that DDM is a facile and robust method for sizing suspensions with multimodal PSDs, but must be coupled with a suitable method for deducing diffusivity distributions from measured ISFs. Specifically, we have demonstrated the use of CONTIN, which is already familiar from long use in DLS. In future work, more advanced algorithms designed for DLS analysis should be explored for potential improvements in resolution and performance.\cite{RegularizedMaxEntropy, SBLAnalysis, CORENNReview, CORENNWeb} In addition, for accurate uncertainty estimates in fitted parameters, correlations between $g(k, \tau)$ points could be included and a Bayesian fitting algorithm may be advantageous as an alternative to least-squares fits. 

We have used bright-field microscopy, but phase-contrast or fluorescence should also be suitable. The method is extendable to turbid systems if suitable $k$ ranges and $g(k, \tau)$ models are used.\cite{DDMTurbid} However, deviation from standard DDM is strongest at low $k$, which may obviate some of the benefits discussed in Section~\ref{sec:Intro}.

In section~\ref{sec:Bidisperse} we have shown how our analysis breaks down when signal contribution $\lesssim 2\%$ or for size ratio $\approx 1:2$. However, breaking these conditions does not necessarily lead to failure; neither is satisfying them a guarantee of success. For example, both conditions are dependent on polydispersity and overlap of populations in the PSD; separately, the limit of applicability may be improved by reducing the uncertainties in the DICF with larger videos, or increasing the signal to noise ratio $A(q)/B(q)$ -- so could be affected by absolute size. Perhaps more important than enunciating hard-and-fast `limits of applicability' is to note how failure occurs. In the case of a small signal from one population, the spread of the 5 repeats typically becomes very large around the mean (accurate but imprecise), which is easy to identify; however, in the case of comparable sizes, the reported values could be consistent but biased (inaccurate but high precision), which might be mistaken for a good measurement.

Before concluding, we summarise a protocol for sizing multimodal suspensions using DDM. One starts by visually inspecting images of the sample, increasing the magnification from a low value until the first particles become visible. At this point, when the largest particles should be comparable to pixel size, record a set of videos and back out the ISF. A CONTIN analysis may already reveal multimodality, or only show a single large population. Regardless, one would then dilute the suspension (and/or increase the magnification) until the signal from the largest particles disappears to reveal smaller populations. If no signal remains, indicated by a time-independent DICF, one may be reasonably confident any small particles are at a number density comparable to or lower than that of the large particles.

We conclude that DDM can fill an important gap between low-throughput electron microscopy and high-throughput DLS. While DLS can access shorter timescales and is likely more sensitive,\cite{Multiscale_Review} form factor effects make multimodal systems challenging. In contrast, access to real space images and low-$k$ information makes DDM uniquely suited for sizing multimodal suspensions, which are ubiquitous in applications.

\section*{Author Contributions}
{\bf Bradley}: Investigation, Methodology, Formal Analysis, Software, Visualization, Writing -- original draft. {\bf Martinez}: Conceptualisation, Methodology, Investigation, Supervision, Formal Analysis. {\bf Arlt}: Conceptualisation, Formal analysis, Methodology, Software, Supervision. {\bf Royer}: Supervision. {\bf Poon}: Conceptualisation, Formal Analysis, Funding Acquisition, Supervision. {\bf All authors}: Writing -- review and editing. 

\section*{Conflicts of interest}
There are no conflicts to declare.

\section*{Data Availability}
Data relevant to this publication is available at \url{https://doi.org/10.7488/ds/3851}.

\section*{Acknowledgements}

JJB was funded by the EPSRC SOFI$^2$ CDT (EP/S023631/1) and Solvay. VAM, JA and WCKP were funded by ERC PoC award GA 882559 NoChaPFI.




\balance


\bibliography{MultimodalDDM} 
\bibliographystyle{rsc} 

\clearpage
\appendix

\section{DDM -- Scattering Perspective} \label{appendix:ScatteringDDM}
The first demonstration of DDM particle sizing\cite{Cerbino2008DDM} used bright-field imaging with a condenser lens of numerical aperture NA~=~0.9 and an objective of NA~=~0.5. In a follow-up publication,\cite{Giavazzi2009DDM} it was stated that NA of the condenser was `about ten times smaller than the numerical aperture of the objective'. We take this to mean that an iris restricts the actual condenser NA to be around 0.05,  giving partially coherent incident light. Consistent with this, the authors analyse their experiment in terms of the interference of the electric fields of light that has and has not been scattered by particles in the sample, treating their technique as a development of dynamic heterodyne near field scattering (HNFS).\cite{DynamicHNFS,HNFS_Ferri_2004} Again, consistent with this understanding, a subsequent review\cite{giavazzi_digital_2014}  shows (their Fig.~3b) that their raw images consist of speckle (= random interference) patterns.

Other researchers have been able obtain results without stopping down the condenser. Thus, our bright-field imaging uses a condenser iris with NA an order of magnitude larger than in the original implementation; others have implemented DDM using fluorescence or absorbance to generate contrast. In all of these cases, there is no coherence in the incident light, but DDM still extracts the ISF. This is because, generally, linear space-invariant images can be treated as density maps, so that correlating images directly gives density autocorrelations. Whilst there are subtle differences, the overall outcome remains the same, with the notable exception that scattering requires a small correction to $k$ for phase shifts due to motion perpendicular to the image plane.\cite{Giavazzi2009DDM}
 
In section~\ref{sec:PolydisperseDDM} we show theoretically that $A(k) \sim NR^6$. Our derivation implicitly assumes that the light is incoherent, as we only consider intensities: the overall intensity contrast of a single particle is calculated by integrating over the particle's volume, where each volume element contributes with the same intensity contrast density $\rho$. The correctness of this result for small to moderately large particles is then demonstrated experimentally in section~\ref{sec:Scaling} for brightfield DDM with incoherent illumination.

In the heterodyne scattering mode of DDM, the fluctuations in the camera signal will be proportional to the scattering field.\cite{giavazzi_digital_2014, Multiscale_Review} The scattering of light by particles is well understood in terms of size and shape\cite{BombannesPusey, LS_Small}; in particular, the scattered field will scale with $V_p \sim R^3$ for sufficiently small (Rayleigh) scatterers, therefore from a heterodyne perspective we can obtain the same $A(k) \sim NR^6$ scaling when calculating the correlation function. Note that for larger particles, such as those typically sized using DDM (including this work), the size scaling of the scattered field is no longer trivial, as the Mie scattering pattern shows strong directional dependence.
\vspace{3cm} 

\section{Example DICFs} \label{appendix:DICF_examples}
\begin{figure}[b!]
\centering
  \includegraphics[width=\linewidth]{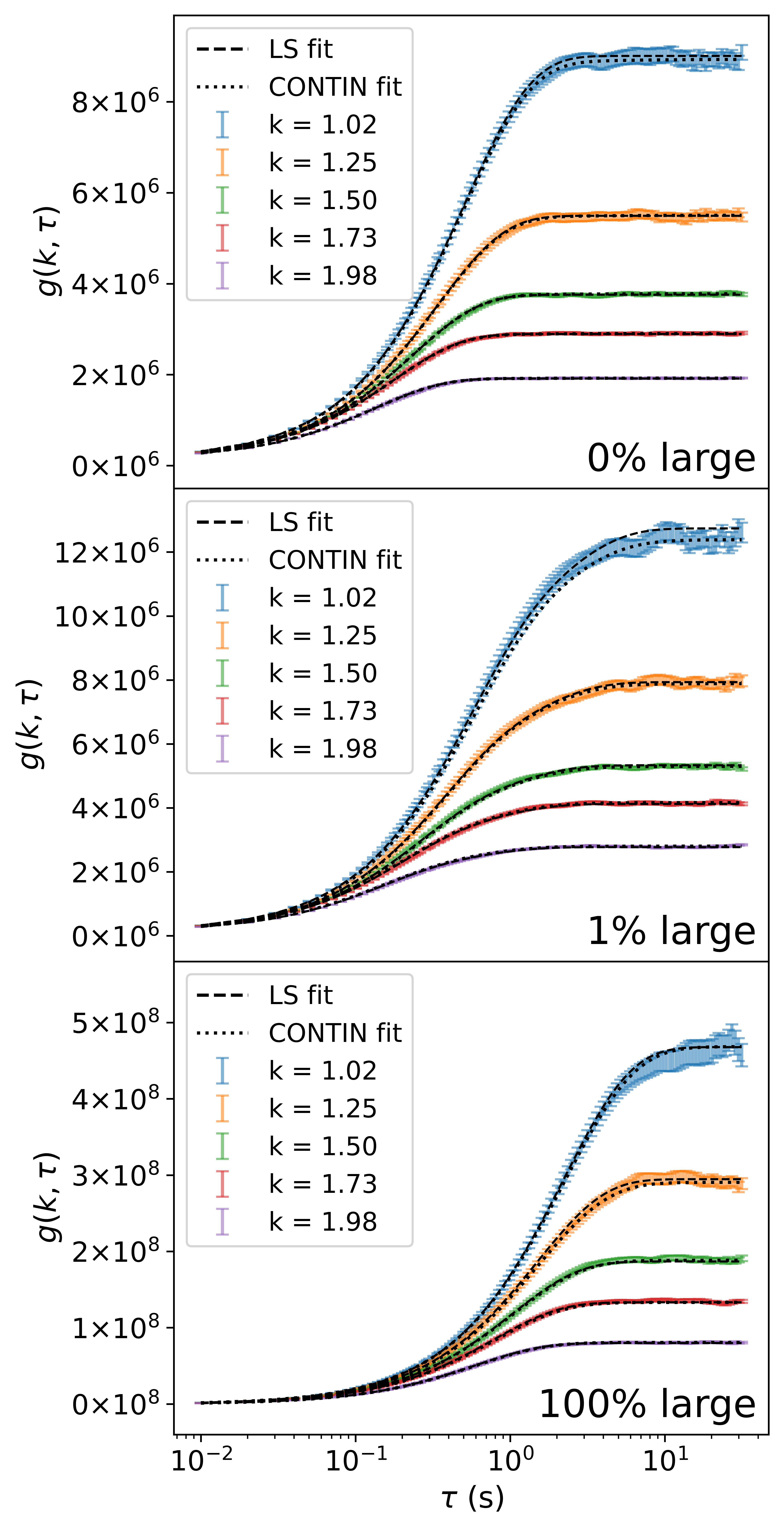}
  \caption{Representative differential image correlation functions, $g(k, \tau)$, at selected $k$ values for the monomodal and bimodal suspensions discussed in section~\ref{sec:Bidisperse}, along with the corresponding least-squares (LS) and CONTIN fits. The top and bottom frames show the signal from suspensions of \SI{240}{\nano\meter} and \SI{1.1}{\micro\meter} respectively, and the middle frame corresponds to a mixture which is 1\% large particles by mass (contributing $\approx 37\%$ of the signal).}
  \label{fgr:Factor_4_DICFs}
\end{figure}

Figure~\ref{fgr:Factor_4_DICFs} shows the extracted DDM signals (in the form of the DICF, $g(k, \tau)$) for two monomodal suspensions and a mixture of the pair. By eye, it is not obvious that the DICF for the bimodal suspension is not monomodal, however by fitting appropriate models it is possible to extract both sizes and their relative contribution to the signal (see section~\ref{sec:Bidisperse}).

\section{DDM with Multiple Populations} \label{appendix:DDM_Theory}
The derivation of Wilson \textit{et al.}\cite{WilsonPRLDDM} assumes that intensity fluctuations are proportional to density fluctuations. However, signal from a given population depends on particle size as well as concentration (see Sections~\ref{sec:PolydisperseDDM} and \ref{sec:Scaling}). Therefore, this assumption does not hold outside of monodisperse suspensions. Instead, consider $M$ independent populations of particles, where overall fluctuations in intensity are given by additive components proportional to fluctuations in density of each population; 
\begin{equation} \label{eq:proportional_fluctuations}
\Delta I (\vec{r}, \tau) = \sum_{i}^M \kappa_i \Delta \rho _i (\vec{r}, \tau)
\end{equation}
where $\Delta \rho_i = \rho_i (\vec{r}, \tau) - \left< \rho_i \right>$ are the fluctuations in the density $\rho_i$ of population $i$. The constants $\kappa_i$ will depend on imaging setup and particle properties which determine contrast -- such as refractive index, size, and physical image formation mechanisms. The Fourier transformed difference images are then given by 
\begin{equation}\label{eq:FT_Diff}
D(\vec{k}, \tau) = \sum_{i}^M \kappa_i\left[\Delta \rho_i (\vec{k}, \tau) - \Delta \rho_i (\vec{k}, 0) \right].
\end{equation}
We can then invoke the key assumption that the populations are independent, so density fluctuations of each are uncorrelated. In this case, when taking the time averaged square modulus of equation \ref{eq:FT_Diff} to obtain the DICF, all components of the inter-population cross-terms will be zero. The remaining terms can be written
\begin{equation} \label{eq:DICF_terms}
g(k, \tau) = \sum_{i}^M 2 \kappa_i^2 \left< \left| \rho_i (k) \right| ^2 \right> \left[ 1 - \frac{\left<\Delta \rho_i (k, \tau) \Delta \rho_i^*(k, 0)\right>_t}{\left< \left| \rho_i (k) \right| ^2 \right>}\right],
\end{equation}
where $\left< \left| \rho_i (k) \right| ^2 \right>$ describes the sample structure. The final term in the summand of Eq.~\ref{eq:DICF_terms} can be recognised as the ISF for the $i$th population $f_i (k, \tau)$, so by defining $A_i (k) = 2 \kappa_i^2 \left< \left| \rho_i (k) \right| ^2 \right>$ we can write the entire multi-population DICF as 
\begin{equation}\label{eq:two_pop_DICF}
    g(k, \tau) = \sum_i^M A_i(k) \left[ 1 - f_i (k, \tau) \right]
\end{equation}
An additive noise term $B(k)$ can also be included for instrument noise. Equation \ref{eq:two_pop_DICF} shows that the contributions to the DICF are addditive, by defining 
$A(k) = \sum_i^M A_i(k)$ and $C_i(k) = A_i(k)/A (k)$, so $\sum_i C_i(k) = 1$, we can write this in the more familiar form
\begin{gather} 
g(k, \tau) = A(k) \left[1 - f(k, \tau)\right] + B(k),\label{eq:ap_DICF} \\
\textrm{where } f(k, \tau) = \sum_i^M C_i(k) f_i(k, \tau). \label{eq:final_two_pop_ISF}
\end{gather}
Equation \ref{eq:ap_DICF} is the usual expression for $g(k, \tau)$ which is key to DDM analysis, with an ISF which is the weighted sum of the ISFs of the individual populations.

The full theoretical calculation of population signal strength $A_i(k) \propto \kappa_i^2 \left< \left| \rho_i (k) \right| ^2 \right>$ is challenging, although fortunately not a requirement to use DDM.  

\vspace{0.5cm}

\section{Cumulant fit to bidisperse measurements}\label{appendix:Cumulant_Fits}
\begin{figure}[h!]
\centering
  \includegraphics[width=\linewidth]{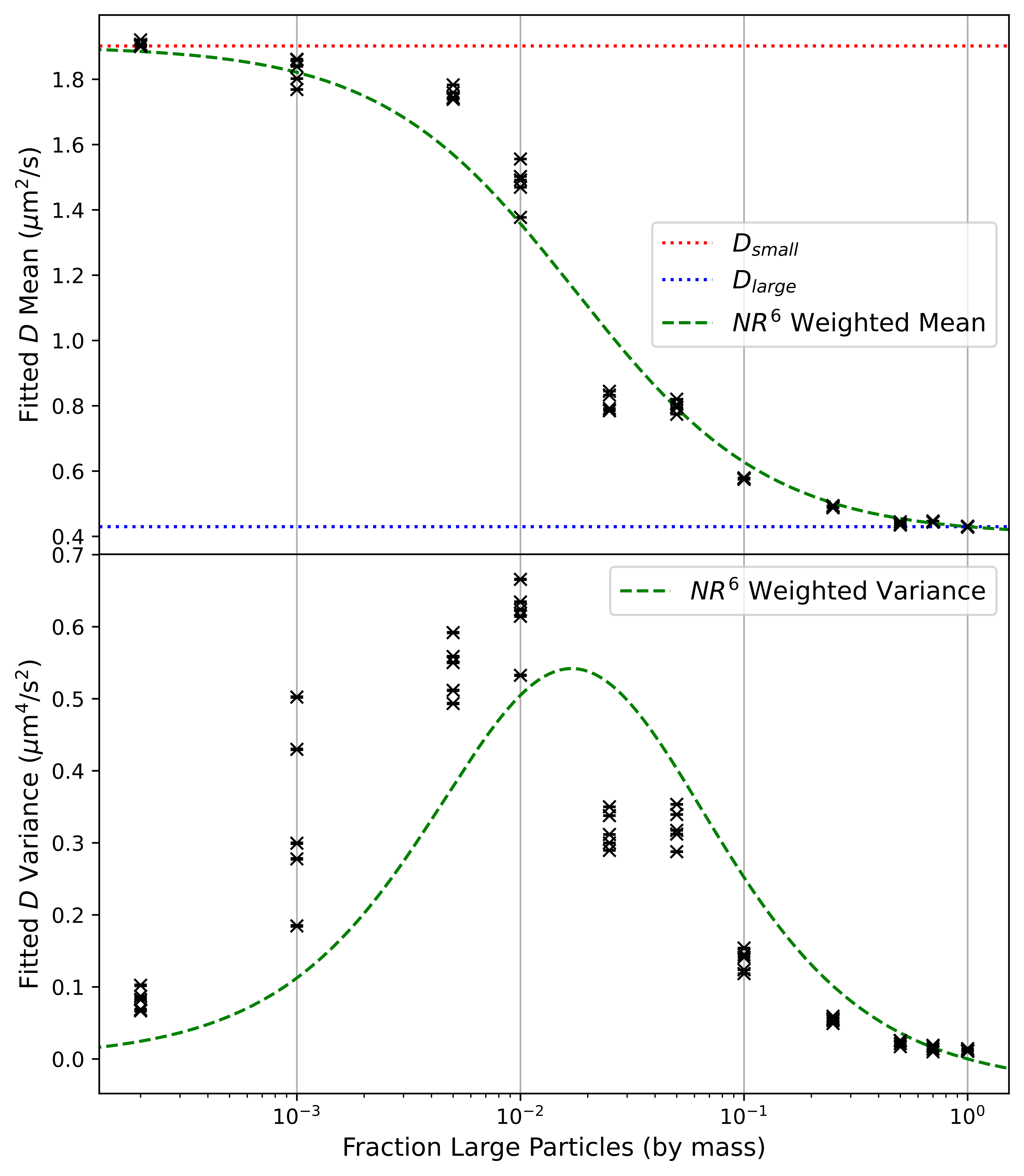}
  \caption{Fitted PDD moments for the bimodal mixtures described in Table~\ref{tbl:bidisperse_mix} (section~\ref{sec:Bidisperse}), obtained using 3rd order cumulant fits to the ISF. Each data point is the result from fits to a single video, with corresponding uncertainties as error bars. Top~-- mean diffusion coefficients, the dotted horizontal lines show the average fitted diffusion coefficient for the monomodal suspensions, the dashed green line shows the mean diffusion coefficient, weighted by the expected signal contributions. Bottom -- corresponding variance from the same fits. Dashed green line shows the weighted variance predicted (assuming monodispersity of each population).}
  \label{fgr:Factor_4_Cumulant}
\end{figure}

Figure \ref{fgr:Bidisperse_LS_fig} shows least-squares results from fitting an explicit bidisperse model (model~3 in Section~\ref{sec:leastsqintro}) to bimodal suspensions with compositions in Table~\ref{tbl:bidisperse_mix}, obtaining good results for each particle size. If we did not know {\it a priori} of bimodality, we may instead fit model~2 in Section~\ref{sec:leastsqintro},
Fig.~\ref{fgr:Factor_4_Cumulant}. Such a cumulant fit to the same data extracts moments comparable with the Eq.~\ref{eq:R3scaling} weighted mean and variance of the diffusion coefficient distribution; in other words, the fit returns reasonable first and second moments of the PDD. The reported variance is slightly less accurate than the mean, which is expected as the accuracy of terms in cumulant fits is known to decrease with increasing order,\cite{2015Cumulant} and the cumulant is designed for reasonably narrow distributions where a continuous curvature is observed at the mean, which is not generally true for a bimodal system. Nevertheless, the results are reasonable, although do not obviously indicate the model's inappropriateness for a bidisperse system.

\vspace{2cm}

\section{Extracting Composition of the Trimodal Suspension}\label{appendix:ROI_composition}
In Section~\ref{sec:Trimodal} we showed that by selecting subregions of an image to analyse it is possible to identify three distinct populations in the PDD of a trimodal suspension. However, these analyses are performed independently, so the relative contributions of each population to the signal is not extracted directly in a single histogram. Here we outline a procedure by which the relative volume fractions of the populations can be extracted from the CONTIN analyses.

The full video contains signal from all particles, and the selected 128\x{}128 ROI only contains signal from the two smaller populations. Therefore, the fraction of the signal from the two smaller populations is ${A_{\rm ROI}(k)}/{A_{\rm full}(k)}$, and the fraction from large particles alone can be obtained as
\begin{equation}
    C_{\rm large} = 1 - \left<\frac{A_{\rm ROI}(k)}{A_{\rm full}(k)}\right>_k
\end{equation}
where $A_{\rm full}(k)$ and $A_{\rm ROI}(k)$ are the (size normalised) DDM signal strength $A(k)$ for the full video and smaller subregion respectively. We average this ratio over the range of $k$ which overlap in the two analyses.

The relative contributions of the small and medium particles to the signal in the reduced video can be extracted from the CONTIN result by summing the bins contributing to each peak (as noted in the main text, due to quadrature weight settings it is the sum of bin heights which determines contribution in this case, not the area). This fraction is labelled $c_{j}$. The fractional contribution of the small/intermediate population $j$ to the total signal is therefore 
\begin{equation}
    C_{j} = c_j \left<\frac{A_{\rm ROI}}{A_{\rm full}(k)}\right>_k.
\end{equation}
Equation~\ref{eq:R3scaling} allows signal strength to be related to volume fractions. In these videos we consider a smaller region at high magnification and therefore inevitably work at a relatively high $k$; for the large particles in particular the impact of the form factor is not negligible even though we remain far from the minimum so that no populations are lost. Volume fractions of the three populations are given by
\begin{equation}
    \phi_i = \gamma \frac{C_i}{R_i^3 \left<P(kR_i)\right>_k^2},
\end{equation}
where again we average the form factor $P(kR)$ over the range of $k$ used and $\gamma$ is a constant of proportionality. When considering relative volume fractions $\phi_i/\sum_k \phi_k$, the value of $\gamma$ is irrelevant.

This can be done for each of the 5 pairs of videos (full video and subregion) to obtain a measure of uncertainty; for each $\phi_i$ five values will be obtained. The mean can then be quoted along with the standard error ($\sigma/\sqrt{5}$) for final values and uncertainties which we find are in very good agreement with the true composition of the trimodal mixture used for these experiments.
\vspace{3cm}

\section{Additional Data and Plots}\label{appendix:Full_Data}

\subsection{Least-squares and CONTIN fits to 240~nm/500~nm videos}\label{subapp:Factor_2_res}
\begin{figure}[h!]
\centering
  \includegraphics[width=\linewidth]{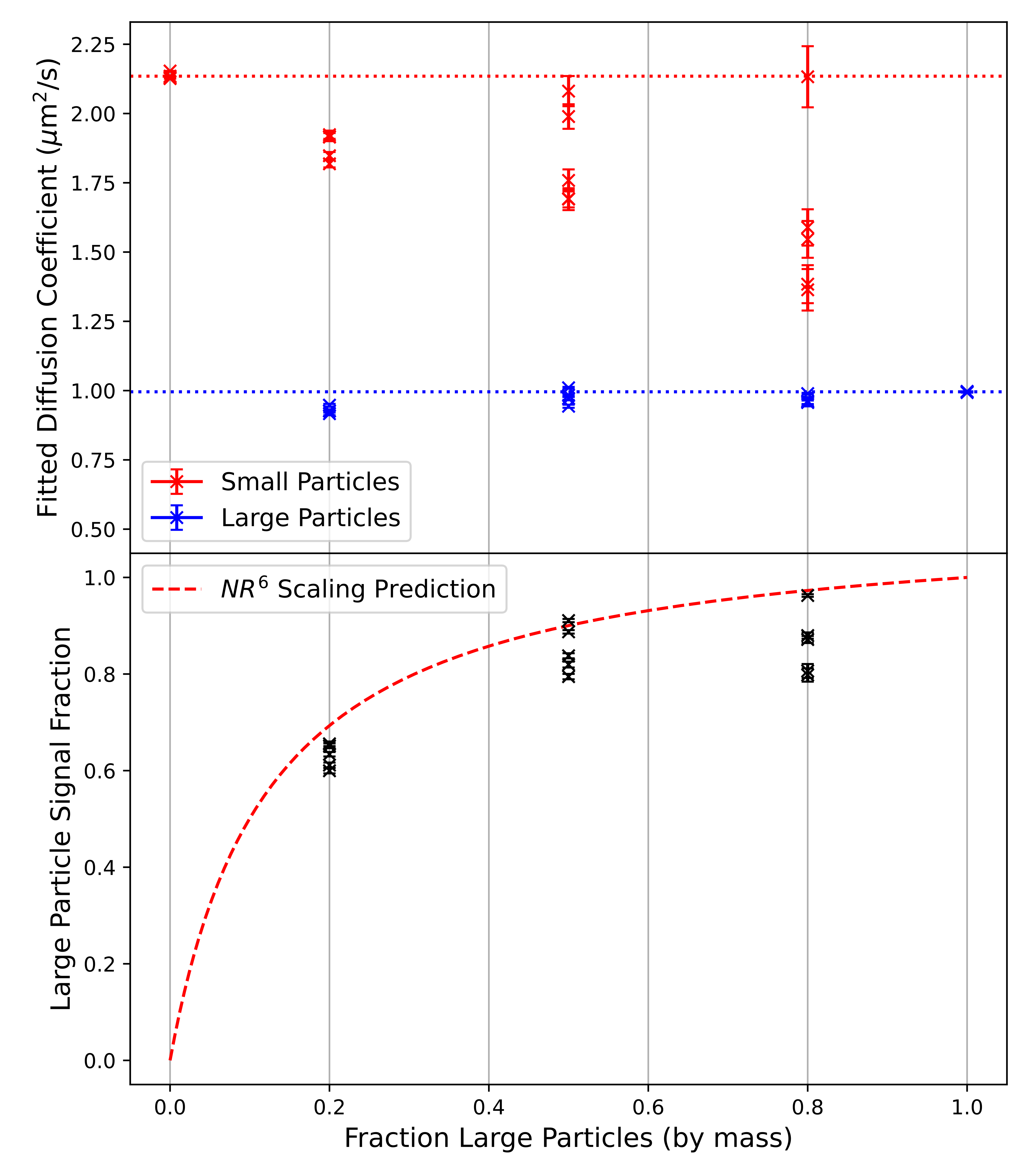}
  \caption{Parameters obtained from bidisperse least squares fits of the DICFs from 4:1, 1:1, and 1:4 mixtures (by mass) of 240~nm and 500~nm particles. Each data point is the result from fits to a single video, with corresponding uncertainties as error bars. a) Fitted diffusion coefficients, the dotted horizontal lines show the average fitted diffusion coefficient for the monomodal suspensions. b) Fitted signal fraction from the large particles, with a prediction based on the $NR^6$ scaling of Eq.~\ref{eq:R3scaling}.}
  \label{fgr:Factor_2_LS}
\end{figure}

Bidisperse mixtures of 240~nm and 500~nm particles were prepared by mixing mass fraction $10^{-4}$ suspensions of monodisperse particles, with ratios of 4:1, 1:1, and 1:4. Large particles are expected to contribute 71\%, 91\%, and 97\% of the signal respectively, based on eq.~\ref{eq:R3scaling}. These were recorded at 200~fps with 20\x{} magnification and no binning (pixel size \SI{325}{\nano\meter}).
They were then fitted to the DICF with a bidisperse model (see Section~\ref{sec:Bidisperse} of the main text) to obtain a mean diffusivity for each population and the relative contribution to the signal, Fig.~\ref{fgr:Factor_2_LS}. We see that in the five repeats the bimodal suspension shows significant variability, switching between accurate results and underestimated values for both populations. This grouping could be an indication of local minima in $\chi^2$.

\begin{figure}[h!]
\centering
  \includegraphics[width=\linewidth]{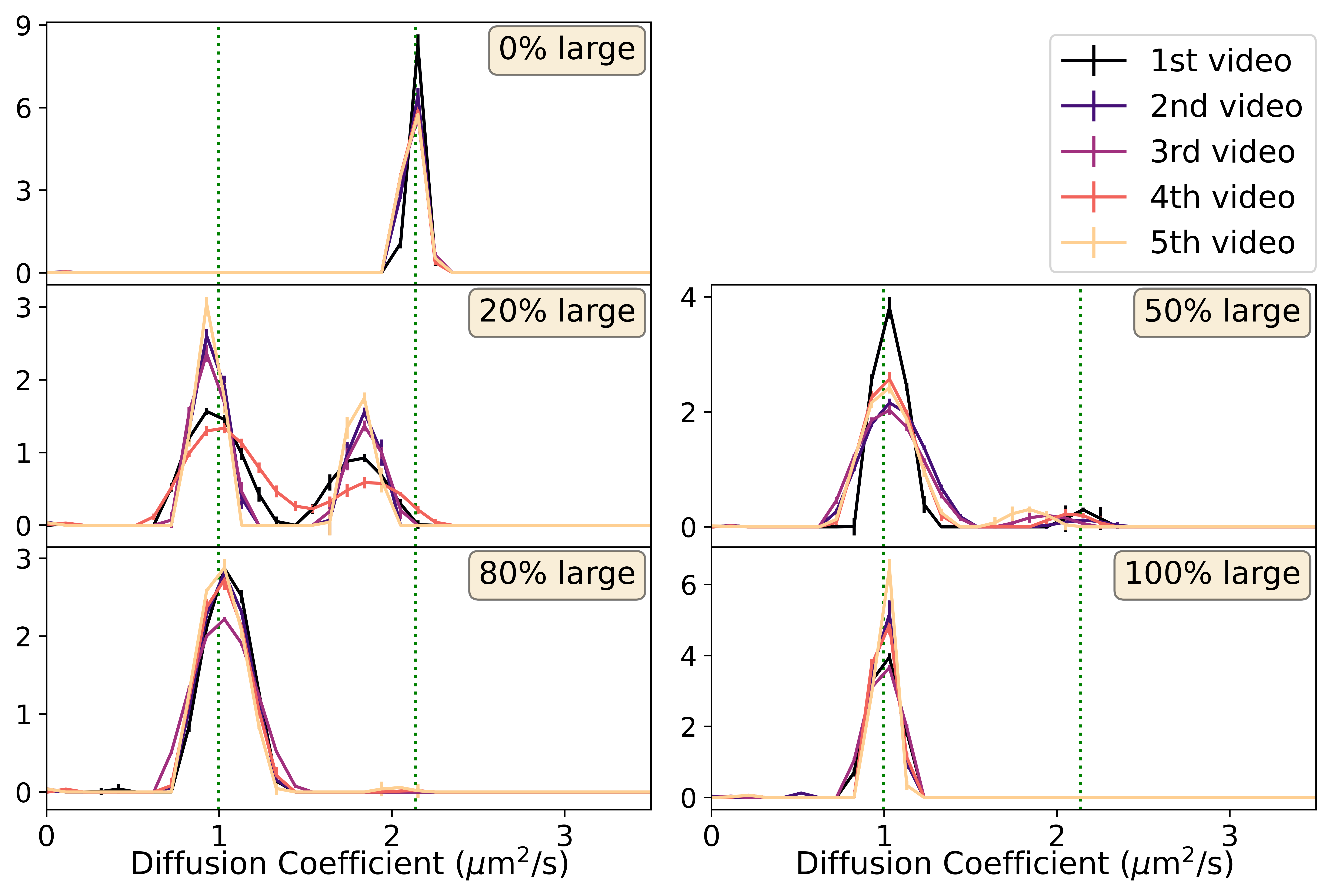}
  \caption{CONTIN results for mixtures of 240~nm and 500~nm spheres mixed in different proportions. Each plot is labelled and had identical analyses performed, the vertical dotted lines show the average diffusion coefficient fit to videos of the separate populations.}
  \label{fgr:CONTIN_factor_two}
\end{figure}

CONTIN fits were performed to the same videos, Fig.~\ref{fgr:CONTIN_factor_two}. Again we see two populations when the contribution of the minority signal component $\gtrsim 5\%$, with diffusion coefficients comparable to the values obtained by LS fits (Fig.~\ref{fgr:Factor_2_LS}).

\subsection{All CONTIN results from ROI selection (Section~\ref{sec:Trimodal})}\label{subapp:ROI_Full}
\begin{figure}[h!]
\centering
  \includegraphics[width=\linewidth]{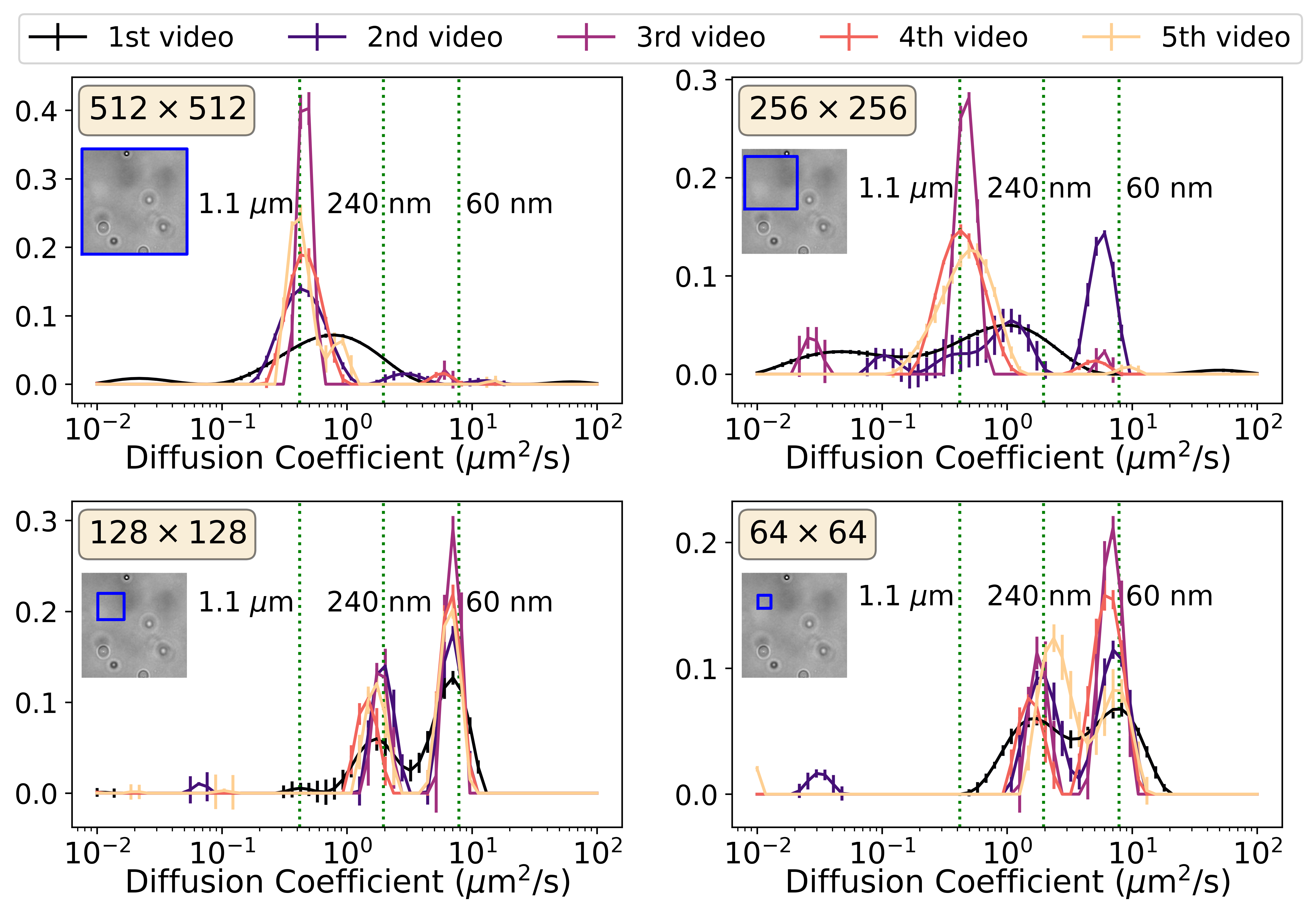}
  \caption{CONTIN fits applied to sub-regions of the 60\x{} magnification videos of the original stock mixture from Table \ref{tbl:trimodal_mix}. Each plot is labelled with the size of the ROI, which is illustrated for the first video in each set by the blue box in the inset. Vertical dotted lines show expected peak positions, labelled with particle diameter.}
  \label{fgr:ROI_Full}
\end{figure}

Figure~\ref{fgr:ROI_Full} shows all results for the ROI selection to exclude large particles from a trimodal suspension. An example for a single video with full and 128\x{}128 ROI was shown in Fig.~\ref{fgr:Trimodal_ROI} of Section~\ref{sec:Trimodal}. In Fig.~\ref{fgr:ROI_Full} we plot all 5 repeats for each video and selected subregions of various sizes. As the ROI is reduced, the largest particles are removed with increasing levels of success, until at 128\x{}128 pixels the two smaller populations are clearly visible in all repeats with no contribution from the large particles. As with the bimodal CONTIN experiments, we see significant variation in shape between repeats, although the overall trend is clear.

\topfigrule

\subsection{Complete CONTIN results for 240~nm/1.1\micron{} videos}\label{subapp:Factor_4_CONTIN}
\begin{figure*}[b]
\centering
  \includegraphics[width=\linewidth]{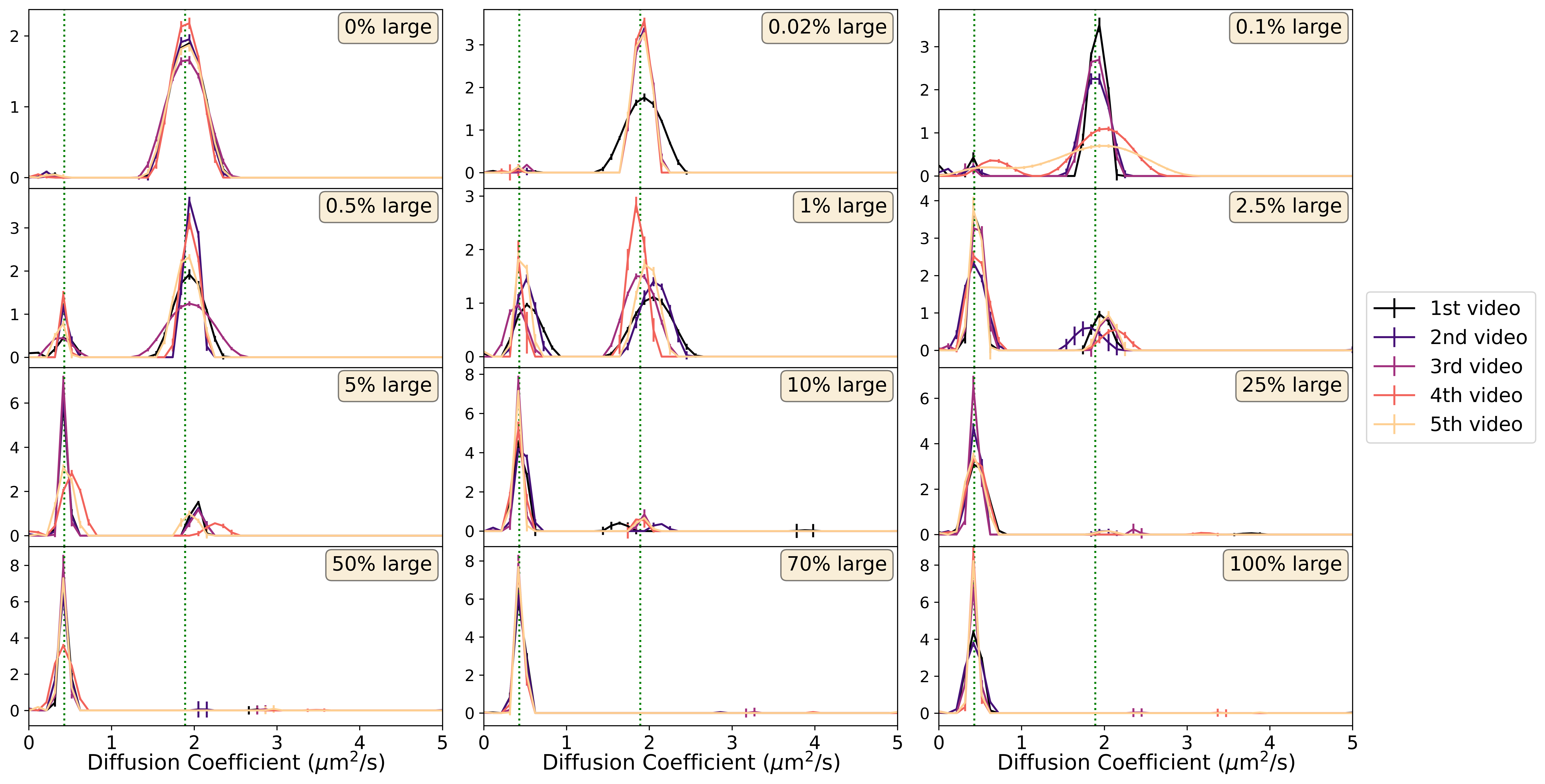}
  \caption{Complete CONTIN results for mixtures of 240~nm and 1.1\micron{} spheres mixed in different proportions listed in Table~\ref{tbl:bidisperse_mix} (Section~\ref{sec:Bidisperse}). Each PDD in each plot is from a single video and had identical analyses performed. The vertical dotted lines show the average diffusion coefficient from least-squares fits to videos of the separate populations.}
  \label{fgr:CONTIN_Bidisperse_Full}
\end{figure*}

Figure~\ref{fgr:CONTIN_Bidisperse_Full} shows the results for CONTIN fits to every one of the videos recorded to produce Figs.~\ref{fgr:Bidisperse_LS_fig} and~\ref{fgr:CONTIN_Bidisperse} in Section~\ref{sec:Bidisperse}. The CONTIN particle diffusivity distributions (PDDs) agree reasonably well with the least-squares fits as discussed in the main text, and as long as the signal from a population exceeds 5\% the peak is consistently present in the PDD, and as the signal drops towards this limit the position and presence of the peak becomes much less reliable. In addition, we can see variation in the width of the peaks between repeats, which we attribute to variability in the selection of the regularisation parameter $\alpha$. This could potentially be eliminated in certain situations (e.g. repeatedly sizing a suspension for consistency as part of quality control) by selecting a fixed~$\alpha$ for the analysis.

\vspace{17 cm}

\end{document}